\documentclass[preprint,12pt]{elsarticle}

\usepackage{amssymb}
\usepackage{amsmath}
\usepackage{array}
\usepackage[table,xcdraw]{xcolor}
\usepackage{tabularx}
\usepackage{colortbl}
\usepackage{booktabs}
\usepackage{graphicx}
\usepackage{float}
\usepackage{footnote}
\usepackage{rotating}
\usepackage{multirow}
\makesavenoteenv{tabularx}
\makesavenoteenv{table}
\usepackage{algorithm}
\usepackage{algpseudocode}
\usepackage{tikz}
\usetikzlibrary{shapes, arrows, positioning, fit, shapes.geometric}
\usepackage[hyphens]{url}
\usepackage[hidelinks]{hyperref}
\usepackage{amsmath}
\usepackage{enumitem}
\usepackage{inconsolata}
\usepackage{listings}
\usepackage[utf8]{inputenc}   
\usepackage{booktabs,tabularx,makecell}
\newcolumntype{Y}{>{\centering\arraybackslash}X}
\newcolumntype{L}[1]{>{\hsize=#1\hsize\raggedright\arraybackslash}X}
\usepackage{natbib}
\biboptions{sort&compress}

\definecolor{niceblue}{HTML}{0000FF}

\usepackage[hidelinks]{hyperref}

\begin{document}

\begin{frontmatter}



\title{Minimal Data, Maximum Clarity: \\A Heuristic for Explaining Optimization}


\author[ncsu]{Amirali Rayegan} 
\author[ncsu]{Tim Menzies} 

\affiliation[ncsu]{organization={Department of Computer Science, College of Engineering, North Carolina State University},
            addressline={890 Oval Drive}, 
            city={Raleigh},
            postcode={27695}, 
            state={North Carolina},
            country={USA}}

\begin{abstract}
Efficient, interpretable optimization is a critical but underexplored challenge in software engineering, where practitioners routinely face vast configuration spaces and costly, error-prone labeling processes. This paper introduces EZR, a novel and modular framework for multi-objective optimization that unifies active sampling, learning, and explanation within a single, lightweight pipeline. Departing from conventional wisdom, our Maximum Clarity Heuristic demonstrates that using less (but more informative) data can yield optimization models that are both effective and deeply understandable. EZR employs an active learning strategy based on Naive Bayes sampling to efficiently identify high-quality configurations with a fraction of the labels required by fully supervised approaches. It then distills optimization logic into concise decision trees, offering transparent, actionable explanations for both global and local decision-making. Extensive experiments across 60 real-world datasets establish that EZR reliably achieves over 90\% of the best-known optimization performance in most cases, while providing clear, cohort-based rationales that surpass standard attribution-based explainable AI (XAI) methods (LIME, SHAP, BreakDown) in clarity and utility. These results endorse   “less but better”;   it is both possible and often preferable to use fewer (but more informative) examples to generate label-efficient optimization and explanations in software systems. To support transparency and reproducibility, all code and experimental materials are publicly available at https://github.com/amiiralii/Minimal-Data-Maximum-Clarity.
\end{abstract}


\begin{highlights}
\item \textbf{Maximum clarity heuristic:} better optimization with less but key data
\item \textbf{Introducing EZR:} an interpretable and modular tool for multi-objective optimization
\item \textbf{Comprehensive validation:} On 60 datasets for robust optimization results
\item \textbf{Explanation:} Practical, actionable explanations to guide real-world decisions
\item \textbf{Downstream validation:} showing explanations improve feature selection
\end{highlights}


\end{frontmatter}



\section{Introduction}
This paper addresses the challenge of multi-objective optimization in software engineering (SE), such as configuration tuning and hyperparameter optimization. Selecting optimal configurations is critical for system performance, reliability, and cost-efficiency. However, in practice, optimization is often poorly managed. Software systems expose numerous configuration options, yet practitioners typically rely on defaults or intuition, leading to sub-optimal results and frequent failures \cite{ref13}. Many systems are deployed with generic default configurations that are rarely revisited, even though they may not suit the target environment. Administrators often retain these defaults due to limited time, expertise, or tool support, which can lock systems into suboptimal performance states \cite{ref32}. This gap between potential and practice means that large parts of the configuration space remain unexplored, leaving valuable performance gains untapped and making systems more prone to inefficiencies and hidden defects. In fact, empirical data shows that over 60\% of configuration variables remain unchanged after their initial creation, highlighting how rarely these options are actively optimized \cite{ref31}. This becomes a major pain point because exhaustive evaluation is infeasible, given the vast configuration spaces and the complex, non-linear interactions among parameters \cite{ref12}.

A major bottleneck in this task is the
{\em labeling problem}. While collecting unlabeled configurations is easy, obtaining their labels is expensive, time-consuming, and often noisy. Creating labeled datasets in software engineering is costly and slow, as it requires expert knowledge to interpret complex artifacts like bug reports. For example, running performance tests for all configurations is computationally prohibitive \cite{ref44}. Defect labels, derived heuristically from commit messages,  are error-prone \cite{ref45, ref46}.  Studies show that even carefully labeled datasets often contain errors, and the high cost of manual labeling limits the scale of available data \cite{ref33}.
Therefore, SE optimization methods must:
\begin{itemize}[noitemsep, topsep=0pt]
    \item Minimize labeling cost, using as few labels as possible.
    \item Identify near-optimal configurations efficiently without full Pareto front\footnote{The Pareto front is the set of non-dominated solutions in a multi-objective problem, representing trade-offs where improving one objective requires sacrificing another. Solutions on the Pareto front are all considered optimal.} estimation.
    \item While providing interpretable outputs, enabling practitioners to understand recommendations.
\end{itemize}

To address these goals,  we propose a counterintuitive yet effective heuristic:

\begin{center}
{\em The     Maximum Clarity Heuristic}:\\ To explain complex tasks, \textbf{use less data}.
\end{center}

Consider decision tree learning. Trained on tens of thousands of examples, such trees often span many pages, making them difficult to read and even harder to reason about. But suppose we can identify just 32 highly representative exemplars from the same data. Training the same decision tree learner on that subset (with a minimum leaf size of 2) yields a tree of no more than 16 lines, referencing at most 8 variables (given binary splits). This concise structure can often communicate core decision logic far more effectively than a sprawling tree trained on the full dataset. Figure \ref{fig:sample-ezr-tree} shows the result of our trials toward this goal. It splits the whole data space into handful representative leaves. The column $n$ shows the number of data samples on each branch and the column $win$ shows the winning rate of each branch, stating how good that branch is, based on our goal and our metric which is described thoroughly in the next sections. Lines with the $;$ Symbol at the end of them are the leaves of the tree.

\begin{figure}[ht]
\centering
\small
\begin{lstlisting}
  win    n
----- ----
   15   32
   45   22   feature_1 <= 5:
   20    9   |   feature_2 <= 4:
    5    5   |   |   feature_3  = 'a' ;  (leaf)
   30    4   |   |   feature_3 != 'a' ;  (leaf)
   80   13   |   feature_2  > 4:
   75    7   |   |   feature_4 <= 100 ;  (leaf)
   86    6   |   |   feature_4  > 100 ;  (leaf)
  -10   10   feature_1  > 5: 
   -5    5   |   feature_4 <= 50      ;  (leaf)
  -20    5   |   feature_4 <= 50      ;  (leaf)
\end{lstlisting}
\caption{  EZR generates a very small tree summarizing the major points of the data.  Leaf labels denote cohorts. For a larger example, see Figure \ref{fig:ezr-tree-output}}
\label{fig:sample-ezr-tree}
\end{figure}

\begin{figure}[t]
  \centering
  \includegraphics[width=\textwidth]{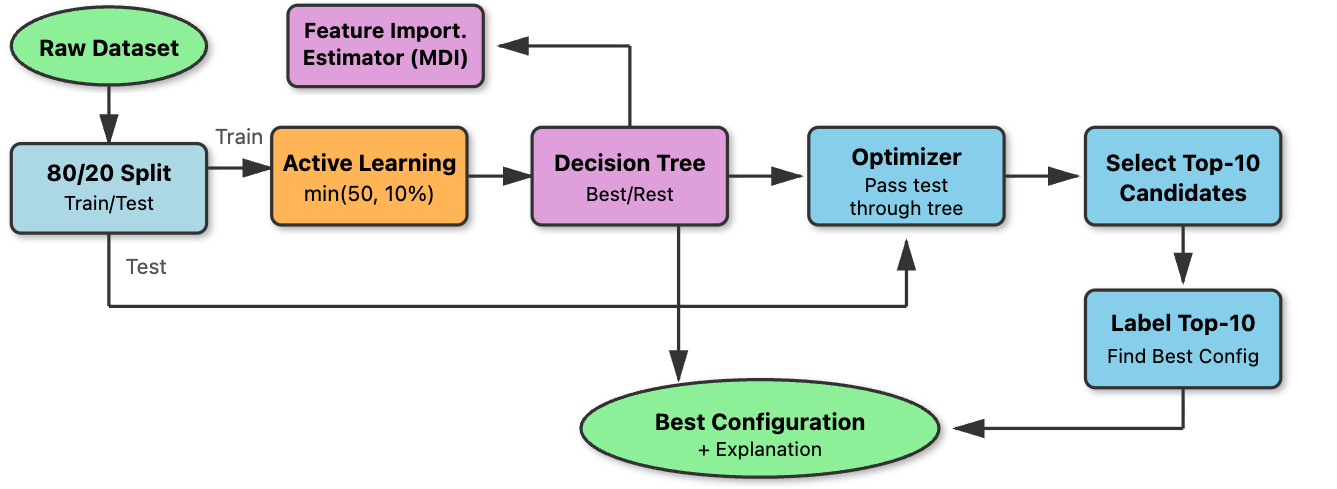}
  \caption{EZR pipeline}
  \label{fig:pipeline}
\end{figure}

We introduce EZR, a novel and modular framework that unifies optimization, learning, and explanation into a single pipeline. As Figure~\ref{fig:pipeline} shows, EZR's source code integrates the following core modules:
\begin{itemize}[noitemsep, topsep=0pt]
\item \textit{Active Learning Sampler:} selects the most informative data points to minimize labeling costs.
\item \textit{Decision Tree Generator:} builds interpretable models that guide structure optimization.
\item \textit{Feature Importance Estimator:} quantifies global contributions of features.
\item \textit{Optimizer:} estimates the likelihood that unlabeled data points achieve optimal multi-objective trade-offs.
\item \textit{Result Explainer:} provides local explanations of why a data point is chosen as the reference optimal and which features drive the decision.
\end{itemize}

While EZR is inspired by reinforcement learning \cite{ref1}, active querying, and Pareto optimization \cite{ref48}, it differs in key ways. Rather than relying on trial-and-error with reward signals, it employs a Naive Bayes sampling strategy that minimizes labeling effort and remains efficient. Unlike traditional active querying, EZR integrates sampling with an interpretable model structure, so each query aids both optimization and explanation. And instead of focusing only on Pareto frontiers, EZR embeds this objective within a modular framework that also supports prediction, feature ranking, and explanation. These distinctions position EZR as a lightweight yet practical alternative for software engineering tasks.

\subsection{Problem Statement and Research Questions}
In this work, we address multi-objective optimization problems in search-based software engineering where each candidate configuration is represented by a feature vector $f=(f_1, f_2, ...,f_n)$ (e.g., configuration options, process parameters). Evaluating a configuration $f$ in the target environment yields multiple objective values (e.g., performance, cost, reliability), but each such evaluation is expensive and often noisy, so only a small labeling budget is available. Our first goal is therefore to design a method that can find high-quality trade-off solutions under a tight label budget, remaining competitive with state-of-the-art optimizers that use many more labeled examples (RQ1). Our second goal is to provide clear, interpretable explanations of these solutions, so that practitioners can understand which features matter most, how changes to them affect the objectives, and why particular configurations are recommended (RQ2).

Even if we achieve label-efficient optimization and good explanations, a further bottleneck remains. There is no widely accepted, practical way to compare different explanation (XAI) methods in this setting. Existing approaches typically rely on proxy metrics or user studies, which are hard to apply systematically across many software engineering optimization tasks. To address this, we propose to compare XAI methods via their impact on downstream optimization performance, using their feature rankings as the basis for feature selection in a separate optimizer, and this motivates our third research question (RQ3). In conclusion, we formulated the following research questions:
\color{black}
 
\begin{itemize}
  \item \textbf{RQ1: Effectiveness on optimization.} To what extent can EZR discover near‐optimal configuration settings, and how does its performance compare to state‐of‐the‐art models? \textit{We show that EZR consistently achieves near-optimal performance ($\ge90$\% of best-known results in 73\% of datasets) while using only a fraction of the labels required by fully supervised models.}
  \item \textbf{RQ2: Comparison to standard XAI.} To what extent can EZR’s “less is more” heuristic generate explanations whose clarity and actionable insight match or exceed those produced by established XAI techniques? \textit{We show that EZR delivers clearer and more actionable explanations than attribution-based methods, spanning all rungs of Pearl’s Ladder of Causation while relying on fewer, more informative features.}
  \item \textbf{RQ3: Practical utility of explanations.} To what extent can EZR’s explanations generate feature rankings that improve downstream optimization performance, and how does this compare to other feature ranking methods?
  \textit{We show that EZR’s explanation-driven feature rankings improve downstream optimization, performing on par with or better than established feature ranking methods while requiring far fewer labels.}
\end{itemize}

All code and experimental materials for this research are publicly available at https://github.com/amiiralii/Minimal-Data-Maximum-Clarity, enabling full reproducibility of the reported results.

The rest of this paper is organized as follows. Section 2 reviews related work and motivates our “less but better” approach with a real‐world case study. Section 3 details the design and implementation of EZR, describes our experimental setup, and summarizes the datasets and evaluation metrics. Section 4 presents a comprehensive evaluation of EZR’s optimization, feature‐selection, and explanation capabilities. Section 5 interprets these results, discusses practical implications and limitations, and lastly, Section 6 concludes with a summary of contributions and directions for future work.

\section{Background}
\subsection{Multi-objective Optimization and Configuration Tuning in SE}

Many problems in software engineering involve balancing competing objectives, such as maximizing performance while minimizing cost, or improving reliability without increasing resource usage. These trade-offs are naturally addressed through multi-objective optimization techniques, which aim to approximate the Pareto front space. In software engineering, multi-objective optimization has been applied to diverse tasks, including software configuration tuning \cite{ref12, ref31}, test case prioritization \cite{ref34,ref35}, requirements selection \cite{ref36, ref37}, and project scheduling \cite{ref38}.

For such multi-objective problems, Search-Based Software Engineering (SBSE) has emerged as a powerful approach for systematically exploring large solution spaces and balancing conflicting objectives in modern software systems \cite{ref42}. SBSE has popularized the use of metaheuristic algorithms (e.g. NSGA-II) for exploring high-dimensional and non-linear optimization spaces \cite{ref38, ref39}. However, these approaches often require large numbers of evaluations, which can be costly in SE contexts where each evaluation may involve building, deploying, and testing complex systems. For example, Nair et al. \cite{ref40} show in the FLASH framework that configuration optimization is often constrained by the high cost of obtaining labeled performance data, and that label-efficient methods can dramatically reduce the number of evaluations needed.

Evaluating candidate solutions in Pareto-based SBSE is   a non-trivial task~\cite{ref41}: 
\begin{itemize}
\item
Recent results at ICSE'26 cast doubts on the value of reasoning over the whole Pareto
frontier. The PROMISETUNE~\cite{ref65} study reports that the
best solutions  are typically clustered into a very small region of the total Pareto  frontier.
This means that traditional assessment methods
(hypervolume, spread, GD, IGD)
are highly misleading since those measures
comment on the whole frontier (and not the tiny regions containing the best solutions).
\item 
While traditional multi-objective optimization methods aim to approximate the entire Pareto front, such approaches require substantial computational resources and large numbers of labeled evaluations to build accurate surrogate models of the search space \cite{ref41, ref40}. 
\begin{itemize}
\item
For instance, methods like NSGA-II and MOEA/D rely on dominance relationships that necessitate pairwise comparisons across the population, while indicator-based approaches require either prior knowledge of the true Pareto front or expensive approximations thereof \cite{ref15}.
\item
Also, another reason not to use standard Pareto-based metrics is that recent work has   demonstrated that computing quality indicators for multi-objective solutions often demands extensive labeled data to establish reference points or ideal vectors \cite{ref57, ref58, ref59}. 
\end{itemize}
\item
In contrast, our goal is fundamentally different. Inspired by the PROMISETUNE
results~\cite{ref65}, rather than modeling the entire solution space or estimating the complete Pareto front, EZR seeks to efficiently identify a small number of high-quality configurations under severe label constraints. This motivates the need for a ranking mechanism that is computationally lightweight (requiring no iterative population updates or dominance sorting) and label-efficient, which operates without access to the full Pareto front.
\end{itemize}
\color{black}
Moreover, recent work has shown that the search spaces of SE problems often contain redundant or weakly relevant dimensions \cite{ref20}, meaning that exhaustive search wastes effort on uninformative configurations. This motivates the need for approaches that can intelligently allocate budget toward the most promising regions of the search space. 

While multi-objective optimization techniques have achieved notable successes in SE, their practical deployment is limited by evaluation cost, noisy objectives, and the difficulty of interpreting why certain solutions are preferred. This creates an opportunity for lightweight, explanation-oriented optimization methods(such as EZR) that can produce competitive results using fewer evaluations while also providing interpretable decision rationales. 

To illustrate a typical optimization scenario in software engineering, consider the \textbf{COC1000} dataset from the MOOT repository\footnote{https://github.com/timm/moot/blob/master/optimize/process/coc1000.csv}. This dataset models a \textit{COCOMO-style} software process, where the objective is to optimize project planning by minimizing several undesirable outcomes. Each row in the dataset represents a hypothetical project configuration described by 20 independent variables, all described in Table \ref{tab:coc1000_features}. The dependent goals include:

\begin{itemize}[noitemsep, topsep=0pt]
  \item \textbf{Risk-}: Project risk (to be minimized)
  \item \textbf{Effort-}: Estimated person-months (to be minimized)
  \item \textbf{AEXP-}: Required analyst experience (to be minimized)
  \item \textbf{PLEx-}: Programming language experience required (to be minimized)
  \item \textbf{LOC+}: Delivered lines of code (to be maximized)
\end{itemize}
The optimization task is to identify configurations that \emph{minimize risks and resource needs while maximizing productivity}. Such trade-offs are typical in software project management, where cost, staffing demands, and output must be balanced. In our experimental setup, the true values of the objectives remain hidden until explicitly queried, and optimization performance is assessed by the actual quality of the selected configurations while keeping the number of queries (i.e., labels) as low as possible. The dataset contains \textbf{1,001 configurations}, which makes exhaustive evaluation impractical. Therefore, optimizers must select rows \textit{intelligently}, incurring a cost for each query (i.e., when the true goal values are revealed). Note that the 20 independent variables in this dataset represent factors that can take managers years to learn to balance effectively \cite{ref43}.


\begin{table}[htbp]
\centering
\scriptsize
\renewcommand{\arraystretch}{1.3}
\setlength{\tabcolsep}{6pt}
\begin{tabularx}{\linewidth}{>{\raggedright\arraybackslash}p{0.9cm} >{\raggedright\arraybackslash}X}
\textbf{Feature} & \textbf{Description} \\
\midrule
\rowcolor[HTML]{EFEFEF} ACAP & Analyst capability: skill level and experience of analysts \\
PCAP & Programmer capability: general skill of software developers \\
\rowcolor[HTML]{EFEFEF} CPLx & Product complexity: structural and algorithmic complexity \\
DATA & Database size: relative size of data in the application \\
\rowcolor[HTML]{EFEFEF} DOCU & Documentation match: completeness and match to lifecycle needs \\
PMAT & Process maturity: software process capability \\
\rowcolor[HTML]{EFEFEF} RELY & Required reliability: criticality and dependability requirements \\
SCED & Schedule constraint: compression or expansion of timeline \\
\rowcolor[HTML]{EFEFEF} SITE & Team distribution: communication overhead across sites \\
STOR & Storage: main memory constraints of the software product \\
\rowcolor[HTML]{EFEFEF} TIME & Execution time constraints: Performance requirements relative to available computing power \\
TOOL & Tool support: availability and integration of development tools \\
\rowcolor[HTML]{EFEFEF} ARCH &  Architecture and Risk Resolution: Quality of architecture and risk resolution \\
FLEx &  Development Flexibility: Degree of flexibility in development processes \\
\rowcolor[HTML]{EFEFEF} LTEx &  Language and Tool Experience: Familiarity with programming languages and tools \\
PCON &  Personnel Continuity: Turnover rate or stability of the project team \\
\rowcolor[HTML]{EFEFEF} PREC &  Precedentedness: Familiarity with the project type \\
PVOL &  Platform Volatility: Frequency of hardware/software changes \\
\rowcolor[HTML]{EFEFEF} RUSE &  Required Reusability: Level of reusability required for components \\
TEAM &  Team Cohesion: How well the team works together \\
\bottomrule
\end{tabularx}
\caption{Independent variables in the \textit{coc1000} dataset}
\label{tab:coc1000_features}
\end{table}

While efficient optimization is essential, it is not sufficient on its own. In practice, optimizers must also justify their choices so that practitioners can trust and act on them, making explainability a critical complement to optimization. This brings us to the role of explainable AI in software engineering.

\subsection{Explainable  AI in SE}
AI and machine learning models are increasingly used in software engineering for tasks such as defect prediction, code recommendation, automated testing, and maintenance. Although these models often achieve strong predictive performance, their complexity produces “black-box” solutions whose inner workings are opaque to practitioners and decision makers. This lack of interpretability poses both practical and ethical challenges. Stakeholders hesitate to rely on AI tools they cannot understand or justify, especially in critical activities like bug fixing, code review, or requirements analysis. As a result, state-of-the-art models are often viewed as unreliable, costly to validate, or risky to deploy, leading to reduced trust, excessive manual oversight, or outright rejection \cite{ref2}. Crucially, many SE tasks demand not only predictions but also actionable explanations. For example, in defect prediction, simply flagging a file as defective is insufficient and developers and managers need to know why the model reached that conclusion to debug, validate, and align decisions with project goals \cite{ref3}. This demand underscores the importance of explanations that are both accurate and actionable. Stakeholders frequently ask questions such as “Why is this file predicted as defective?” or “Why has this task been recommended for maximum development effort?” Addressing such concerns requires explanations that practitioners across roles, developers, managers, analysts, legal teams, and executives, can trust, validate, and act upon \cite{ref4,ref5}. Recent SE research has therefore shifted from coarse, opaque predictions toward models that provide clear and practical guidance. Explanations should be delivered in accessible forms, such as natural language, decision trees, visual highlights, or actionable recommendations, tailored to the needs of different stakeholders \cite{ref4}. To support this, traditional black-box models are increasingly complemented by interpretability techniques. White-box methods expose feature importance in deep learning models, while black-box methods rely on perturbation and attribution strategies to provide model-agnostic explanations at both local and global levels. Importantly, these explanations must remain understandable to non-experts, avoiding overly technical presentations that hinder adoption \cite{ref6}.

Explanations in AI are generally distinguished as global or local. \textit{Global explanations} describe overall model behavior, identifying which features most influence predictions across the dataset. In this study, we employ the permutation importance framework \cite{ref23}, a model-agnostic technique that quantifies each feature’s contribution by permuting its values and measuring the resulting drop in model accuracy. \textit{Local explanations}, in contrast, clarify why a model produced a particular outcome for an individual instance. To evaluate these, we compare EZR against three widely used methods. LIME (Local Interpretable Model-agnostic Explanations) \cite{ref25} builds a sparse linear surrogate around the target instance to highlight the most influential features locally. SHAP (SHapley Additive exPlanations) \cite{ref26} applies cooperative game theory to compute fair, additive attributions that sum to the difference between the model’s prediction and a baseline. BreakDown \cite{ref27} sequentially decomposes a prediction by adding features one at a time in a fixed order, producing a step-by-step narrative of how each feature shifts the output away from the baseline.

The justification for incorporating explanations into AI systems, as Chen et al. \cite{ref8} argue, arises from a fundamental limitation of purely data-driven optimization: For an analytic model to be adopted and create value, it must persuade users of its findings. This necessity follows from two core psychological principles. First, humans have limited cognitive capacity, as noted by Miller’s “magical number seven” \cite{ref9}. Stakeholders cannot reason about hundreds of variables simultaneously. Without simplified narratives, results from high-dimensional models remain cognitively inaccessible. Explanations reduce this complexity, distilling decisions into understandable rules or factors aligned with human reasoning. Second, stakeholders often hold pre-existing beliefs and intuitions about their domain, and they often experience cognitive dissonance when model outputs contradict their prior beliefs \cite{ref54}. Without a compelling rationale, they are more likely to reject the model than update their mental model. Explanations act as persuasive tools, resolving dissonance and enabling acceptance of counterintuitive results.

Beyond these psychological foundations, the rise of opaque “black-box” models has intensified the need for explanation. Three additional arguments reinforce its importance:

\begin{enumerate}[noitemsep, topsep=0pt]
    \item \textbf{Fostering Trust in High-Stakes Decisions:} Explanations transform blind reliance into informed confidence, a necessity in domains where errors carry high costs.
    \item \textbf{Ensuring Ethical and Regulatory Compliance:}  Explanations enable audits for hidden bias and support legal frameworks such as the EU’s “right to explanation” under GDPR \cite{ref10}. Transparency has thus become a core system requirement \cite{ref11}.
    \item \textbf{Supporting Debugging and Discovery:} For developers and scientists, explanations are indispensable for diagnosing unexpected model behavior and for revealing novel patterns that drive new hypotheses.
\end{enumerate}
In sum, explanations have evolved from a psychological aid into a multi-faceted requirement. They build trust, ensure fairness and accountability, and advance scientific understanding. Far from being an optional feature, explanation is now a foundational pillar of modern, responsible AI.

\section{Methods}
\subsection{Proposed Method:  EZR }
This study introduces EZR, a new multi-purpose framework designed to efficiently perform machine learning tasks in software engineering domain.

By combining active learning's label efficiency with decision trees' inherent interpretability, EZR achieves what Menzies~\cite{ref64} characterizes as "shockingly simple" solutions that almost match or exceed the performance of far more complex alternatives.
Recent work on hyperparameter optimization has emphasized the importance of matching algorithmic complexity to problem complexity. Agrawal et al.~\cite{ref62} proposed using intrinsic dimensionality thresholds (I < 5) to select between simple and complex optimizers.
That is, when the problem is intrinsically simple,
then simple optimizers will suffice. 
EZR operationalizes this principle by using minimal labeled data to construct interpretable models that rival fully supervised approaches. This stands in contrast to methods like DEHB~\cite{ref63}, which require thousands of evaluations, or traditional Bayesian optimization with Gaussian Process Models that scale poorly with dataset size~\cite{ref40}. 

EZR builds upon a rich lineage of hyperparameter optimization and explainable AI research, but it departs from its predecessors in fundamental ways. While inspired by the Tree of Parzen Estimators (TPE) framework proposed by Bergstra and Bengio~\cite{ref28}, EZR simplifies the surrogate modeling approach considerably:
\begin{itemize}
\item
TPE constructs full kernel density estimators to model the probability distributions of "good" and "bad" configurations, requiring multiple kernels to estimate both mean and variance of predictions. 
\item In contrast, EZR replaces this complex density estimation with a lightweight Naive Bayes classifier that directly computes likelihood ratios between "best" and "rest" configurations.
\end{itemize}
This architectural choice aligns with recent findings by Lustosa~\cite{ref20,ref61}, who demonstrated that many software engineering optimization problems exhibit surprisingly low intrinsic dimensionality, often reducing datasets with dozens of raw attributes to just a handful of intrinsic dimensions. 

EZR leverages this insight from Lustosa by employing a simpler acquisition strategy that avoids the exploration-exploitation balancing acts typical of reinforcement learning approaches~\cite{ref1}. Rather than dynamically adjusting between exploratory and exploitative phases, EZR uses a greedy elite-search that focuses sampling budget on configurations with the highest likelihood ratios, effectively performing a Hoare-style quickselect~\cite{ref19} over the configuration space.

EZR's design philosophy also extends to its treatment of interpretability. Where established explainable AI methods like LIME, SHAP, and BreakDown provide post-hoc, attribution-based explanations that remain confined to Pearl's associational rung of causation~\cite{ref53}, EZR's decision tree structure natively supports all three rungs: associations (through global feature importance), interventions (through threshold-based what-if analysis), and counterfactuals (through alternative leaf assignments). 

\color{black}

\begin{algorithm}

\small
\caption{EZR Overall Pipeline}
\label{alg:ezr_overview}
\begin{algorithmic}[1]
\Require dataset name $s$
\State $raw \gets \textsc{Data}(\textsc{CSV}(s))$
\State $train, test \gets \textsc{Split}(raw, 0.8)$
\State $budget \gets \min(50,\; |train|/10)$
\State $acq \gets \textsc{Xploit}$
\State $k \gets 10$
\State $Labels \gets \textsc{ActiveLearn}(train,\; acq,\; budget)$
\State $Root \gets \textsc{Tree}(Labels)$
\State $scores \gets \textsc{passToTree}(Root,\; test)$ 
\State $C \gets \textsc{TopK}(scores,\; k)$
\State $bestRow,\ performance \gets \textsc{WIN}\!\left(\min_{r \in C}\ \textsc{D2H}(\textsc{Label}(r))\right)$
\State \Return $(bestRow,\ performance)$
\end{algorithmic}
\end{algorithm}

To provide an overview of the complete workflow, Figure~\ref{fig:pipeline} and Algorithm~\ref{alg:ezr_overview} outline the main stages of EZR. The process begins by splitting the dataset into training and testing subsets (80/20). An active learning stage then selects $\min(50, 10\%\ of\ the\ training)$ samples for labeling using the \textit{xploit} acquisition strategy. The labeled subset is then used to train an interpretable decision tree. Next, each test instance is subsequently passed through the tree to estimate its utility based on the statistical summary of the leaf it falls into. Finally, the top 10 candidates with the highest estimated performance are labeled to identify the true best configuration. This way the final labeling budget would be 60 (or 10\%\ of\ the\ training + 10). The following sections provide detailed explanations of the components outlined in Algorithm~\ref{alg:ezr_overview}, including active learning, decision tree generation, evaluation methods and win-rate calculation, and budget assignment.

\color{black}

\subsubsection{Ranking Configurations}
A critical design decision in EZR is using a single, unified metric to rank configurations for multi-objective optimization. EZR uses distance-to-heaven (d2h), computed as the Euclidean distance between each row’s normalized objective values and an ideal “heaven” point with optimal values for each objective (since objectives are normalized, each heaven coordinate is 0 or 1, depending on whether that objective is minimized or maximized). This choice meets our two needs: it is computationally lightweight and label-efficient, and it avoids Pareto-front estimation. Configurations with low d2h are guaranteed to be near-Pareto-optimal, since any dominated solution must have a larger distance to the ideal point. Equation~\ref{eq:d2h} defines this single-scalar ranking per row, where $m$ is the set of objectives and $o_i$ is the normalized value of objective $i$.

\begin{equation}
\mathrm{d2h}(x)=\sqrt{\sum_{i=1}^{m}\left(o_i(x)-h_i\right)^2}
\qquad\text{where}\qquad
h_i=\begin{cases}
1 & \text{Maximizing } i\\
0 & \text{Minimizing } i\\
\end{cases}
\label{eq:d2h}
\end{equation}

\color{black}
\subsubsection{Active Learning}
EZR frames configuration optimization as an active learning problem. It begins with a small set of randomly labeled configurations to bootstrap an initial model. Then, at each iteration, it trains a two-class Naive Bayes classifier to distinguish between "Best" and "rest" configurations. Here, "Best" refers to the top-performing $\sqrt{N}$ configurations based on their d2h values, while "rest" includes the remaining $N - \sqrt{N}$ configurations. For each unlabeled configuration, EZR calculates the likelihood of it belonging to the "Best" class divided by the likelihood of it belonging to the "rest" class. The configuration with the highest such ratio is selected for labeling next. This process repeats until a predefined labeling budget is exhausted. By prioritizing configurations that are most likely to improve optimization outcomes, EZR minimizes the number of required labels.

\begin{algorithm}[htbp]
\scriptsize
\caption{Active Learning with \textsc{Xploit}}
\label{alg:actlearn-xploit-core}
\begin{algorithmic}[1]
\Function{Like}{$x, \mathrm{model}$}
  \State \Return \textit{Naive-Bayes log-likelihood with prior and likelihoods}
\EndFunction
\Function{ActiveLearn}{$D$}
  \State $n \gets \textbf{4}$  \Comment{Warm Start}
  \State $stop \gets \textbf{50}$  \Comment{Predefined Labeling Budget}
  \State $shuffle$ $D.\mathrm{rows}$
  \State $\mathrm{todo} \gets D.\mathrm{rows}[n{:}]$
  \State $\mathrm{done} \gets D.\mathrm{rows}[{:}n]$
  \State $cut \gets \mathrm{round}(n^{\textbf{0.5}})$
  \State $\mathrm{best} \gets \mathrm{done}[{:}cut]$
  \State$\mathrm{rest} \gets \mathrm{done}[\mathrm{cut}{:}]$
  \While{$n<stop$}
    \State $n \gets n+1$
    \State $cut \gets \mathrm{round}(n^{\textbf{0.5}})$
    \State $h \gets \max\limits_{x \in todo} \frac{{Like}({x,\mathrm{best}})}{{Like}({x,\mathrm{rest}})}$ \Comment{\textsc{Xploit}: ratio $B/R$}
    \State $\mathrm{best} \leftarrow \mathrm{best} \cup \{h\}$
    \If{$|\mathrm{best.rows}|\ge cut$} 
    \State \textit{Move weakest in best} $\rightarrow$ \textit{rest} 
    \State \textit{Pop weakest in rest}
    \EndIf
  \EndWhile
  \State \Return $(\mathrm{best},\mathrm{rest},\mathrm{todo})$
\EndFunction

\end{algorithmic}
\end{algorithm}

\subsubsection{Parzen Tree}
EZR is inspired by TPE, which models good and bad configurations using Parzen estimators (kernel density estimators) to guide sampling decisions. However, instead of constructing full density estimators, EZR simplifies this process by using a Naive Bayes classifier to approximate the likelihood of configurations being "best" versus "rest". After the active learning loop completes, EZR builds a decision tree over the labeled configurations to extract interpretable rules explaining which features drive optimal outcomes. The tree function partitions configurations based on feature splits that minimize the variance of \textit{d2h} within each branch, producing clear and human-readable explanations. This combination of lightweight active learning and post-hoc decision tree analysis allows EZR to efficiently find high-performing configurations with minimal labels while providing insights into the underlying decision rationale.

In addition to its optimization capability, EZR also functions as a feature-importance ranking tool by leveraging its interpretable decision tree. At each internal node, EZR evaluates how well a feature partitions the data by measuring the reduction in \textit{d2h} variance, recorded as the node’s \textit{impurity}. The overall importance of a feature is then obtained by summing its impurity reductions across the tree, directly aligning with the Mean Decrease in Impurity (MDI) framework used in tree-based models such as Random Forests \cite{ref47}. Formally, the importance of feature $f$ is defined as:

\begin{equation}
MDI(f) = \sum_{n \in \mathcal{N}_f} \sum_{c \in C(n)} \frac{|c|}{\sum_{j \in C(n)} |j|} \cdot Imp(c)
\label{eq:mdi}
\end{equation}

where \( \mathcal{N}_f \) is the set of internal nodes where $f$ is used for splitting, $C(n)$ denotes the children of node $n$, and $Imp(c)$ is the standard deviation of \textit{d2h} values within child node $c$. This formulation highlights features that meaningfully reduce performance variability within configuration clusters.

\subsubsection{Compared to other methods}

While certainly inspired by the
literature on reinforcement learning and active learning, EZR is far simpler.
For example: instead of applying multiple kernels to learn the mean and variance of any prediction
(as in Hyperopt \cite{ref48}), EZR just uses a simple nearest-neighbor check against the
centers of best and rest.  Also, instead of balancing exploration and exploitation
(as done in reinforcement learning \cite{ref1})
EZR just uses the above greedy elite-search.

\section{Experimental setup}

In this study, we conducted an extensive set of experiments to evaluate the effectiveness of our proposed method. The experimental design was structured to systematically assess EZR’s performance in optimization and explainability compared to widely used state-of-the-art methods in software engineering.

\subsection{Optimization Performance: Regression-Based Baselines}
\color{black}
The goal of our experiment is to evaluate the effectiveness of EZR as an optimizer. Before comparing EZR directly to traditional optimization algorithms, we benchmark it against regression-based predictors. For each dataset, several regression models were trained to predict the \textit{d2h} values of configurations. Each model then recommended the configuration with the lowest predicted \textit{d2h}, which we evaluated by comparing its true \textit{d2h} value to that of EZR’s recommendation. This allows us to assess how well each approach identifies configurations near the ideal multi-objective trade-offs. Practically, all of our datasets include ground-truth labels (e.g., true risk, cost, or duration), and thus all \textit{d2h} values are available. However, in our experiment we assume these labels are hidden during optimization, revealing them only at the end to validate the quality of the selected configurations.

To ensure a competitive and comprehensive comparison, we utilized five widely used regression methods popular in the software engineering domain. Each method represents a distinct family of regression algorithms, enabling us to benchmark EZR against a diverse set of predictive approaches. Linear Regression(LR) was chosen to represent classic statistical models, while Random Forest(RF) serves as a representative of tree-based ensemble methods. Light Gradient Boosting Machine (LightGBM or LGBM) captures the performance of boosting algorithms, and a simple Artificial Neural Network(ANN) was included to represent deep learning-based regressors. Additionally, Support Vector Regressor(SVR) was selected to represent kernel-based methods that are effective in capturing complex non-linear relationships. As a baseline, we also included Random Selection, which randomly selects a configuration as the optimal setting, to contextualize the performance gains achieved by other methods. In total, this setup yields seven treatments per dataset; EZR, the five regression models, and the random baseline.

\subsection{Optimization Performance: Optimization Baselines}
To position EZR within the broader optimization literature, we compare it against two widely used state-of-the-art optimizers, SMAC (Sequential Model-based Algorithm Configuration) \cite{ref66, ref67} and DEHB (Distributed Evolutionary Hyperband) \cite{ref63}, alongside a random baseline. Both SMAC and DEHB are popular in software engineering and automated machine learning and represent state-of-the-art membership query inference methods for hyperparameter optimization. DEHB combines differential evolution search with Hyperband's adaptive resource allocation, using successive halving to evaluate configurations, and is considered the current state-of-the-art, outperforming BOHB and Hyperopt \cite{ref68}. SMAC utilizes Random Forests to model the objective function, enabling support for categorical parameters, and proposes new solutions by maximizing the Expected Improvement acquisition function. We conducted two controlled experiments with fixed evaluation budgets:
\begin{itemize}
    \item Using 50 labels
    \item Using 200 labels
\end{itemize}
In each setting, all methods (EZR, SMAC, DEHB, and random) were restricted to the same number of objective evaluations to ensure fairness. Experiments were run on all 60 datasets. To reduce randomness effects, each experiment was repeated 20 times per dataset using different random seeds. In each repetition, after exhausting the budget, each optimizer selected the configuration it predicted to be best. We then revealed its true objective values and computed its \textit{d2h} and corresponding \textit{win} score (to be discussed in later sections). Finally, we applied statistical tests to the 20-run distributions to determine whether performance differences between methods were statistically significant.

\color{black}

\subsection{Explainability Performance}
While most prior work in software engineering has evaluated the effectiveness of XAI methods through subjective assessments (which we also did in the section \ref{subjectiveXAI}), such evaluations, though valuable, may overlook more objective measures of explanatory utility \cite{ref2}. To address this limitation, we adopt a complementary approach inspired by recent XAI evaluations in other domains \cite{ref49, ref50, ref51}. To evaluate the effectiveness of EZR as an explanation method, we operationalize the effectiveness of an explanation method by examining whether its feature-importance rankings can improve downstream predictive performance. The intuition is that a meaningful explanation should identify features that are not only interpretable but also predictive of outcomes in downstream optimization tasks. Thus, by comparing feature subsets derived from different explanation or selection methods, we can objectively assess their utility. EZR was benchmarked against three widely used feature ranking techniques in software engineering domain, ReliefF, SHAP, and ANOVA. \textbf{ReliefF} \cite{ref55} is an instance-based method that estimates feature importance by comparing how well features distinguish between neighboring instances of different classes, making it effective for handling feature interactions and multi-class problems. \textbf{SHAP} (SHapley Additive exPlanations) \cite{ref26} computes feature contributions based on Shapley values from cooperative game theory, offering theoretically grounded and consistent importance scores. \textbf{ANOVA} (Analysis of Variance) \cite{ref56} evaluates the statistical significance of differences in feature means across output categories, serving as a simple yet effective filter-based method for dimensionality reduction. As a reference, we also include a \textbf{no-selection} baseline that uses all available features to assess the added value of applying feature selection.
For each dataset, EZR is first applied as a feature selector to determine the number of features, $k$, to retain. Using this same $k$, all other methods are applied to select their respective top-$k$ features. Each resulting feature subset is then used to train the LGBM regression model, and predictive performance is compared across treatments.

\subsection{Evaluation Protocol}

While statistical rank-based methods such as Scott-Knott \cite{ref30}, as used in \cite{ref15,ref16}, offer a way to group treatments by significance, they do not convey how close or far apart the performances of different methods truly are. In our context, where models operate under strict label constraints, it is expected that no method, including EZR, can fully recover all latent information in the data. However, the value lies in how well EZR balances this limitation against labeling cost. To better capture the effectiveness of this trade-off, we introduce a relative performance metric that quantifies how close EZR gets to the best-known solution found by the strongest state-of-the-art method with access to all labels. Specifically, we define a ratio-based metric, inspired by the concept of normalized regret in reinforcement learning(equation \ref{eq:win-function}), to measure EZR’s efficiency relative to fully supervised baselines.

\begin{equation}
\mathrm{Relative\ Score}\ (\%) 
= 100 \times \frac{\operatorname{win}\!\bigl(\mathrm{d2h}_{\mathrm{EZR}}\bigr)}{\operatorname{win}\!\bigl(\mathrm{d2h}_{\mathrm{SOTA}}\bigr)}
\label{eq:relative-score}
\end{equation}

where $win(d2h)$ denotes the normalized improvement score (higher is better). The function $win(x)$ is defined as:

\begin{equation}
\text{win}(x) = 100 \times \left(1 - \frac{x - \text{min}}{\text{median} - \text{min}} \right)
\label{eq:win-function}
\end{equation}

In the context of optimization, $x$ refers to the predicted \textit{d2h} value of a configuration selected by a given model. The value $min$ is the lowest (most desirable) \textit{d2h} observed in the dataset, and $\text{median}$ is the median of all \textit{d2h} values. This function rescales outcomes such that lower \textit{d2h} values (closer to the ideal) receive higher win scores. A relative score close to 100\% indicates that the model achieves performance comparable to the best method, despite using far fewer labels.

\begin{table}[htbp]
\scriptsize
\centering
\setlength{\tabcolsep}{3pt}
\renewcommand{\arraystretch}{1.1}
\begin{tabularx}{\linewidth}{@{} c L{0.6} L{0.7} L{1.6} c @{}} 
\toprule
\#Datasets &
Dataset Type &
File Names &
Primary Objective &
$x/y$ \\
\midrule
25 & Specific Software Configurations & SS-A to SS-X, billing 10k &Optimize software system settings, Runtimes, query times, and usage data from software configured in various manners & 3–88 / 2–3 \\
\midrule
1 & cloud & HSMGP num & Hazardous Software,Management Program data & 14/1 \\
1 & cloud & Apache AllMeasurements & Apache server performance optimization & 9/1 \\
1 & cloud & SQL AllMeasurements & SQL database tuning & 38/1 \\
1 & cloud & X264 AllMeasurements & Video encoding optimization & 16/1 \\
7 & cloud & (rs—sol—wc)* & misc configuration tasks & 6/1 \\
\midrule
2 & Health Data Sets & healthClose Issues12mths & Predict project health and developer activity & 5/3 \\
\midrule
3 & scrum & Scrum1k, Scrum10k, Scrum100k & configurations of the scrum feature model & 124/3 \\
\midrule
8 & Feature Models & FFM-*, FM-* & Optimize number of variables, constraints and Clause/Constraint ratio & 128–1024/3 \\
\midrule
1 & software process model & COC1000 & Optimize risk, effort, analyst experience, programming language experience & 20/5 \\
4 & software process model & POM3 (A–D) & Balancing idle rates, completion rates and overall cost & 9/3 \\
\midrule
4 & flight, ground, Orbital Space Plane software & XOMO (Flight, Ground, OSP) & optimizing risk, effort, defects, and months. & 27/4 \\
\midrule
2 & Misc & auto93, Wine & Misc & 4/3 \\
\midrule
60 & \textbf{Total} & & & \\
\bottomrule
\end{tabularx}
\caption{Summary of datasets used in this study.}
\label{tab:datasets-summary}
\end{table}

\subsection{Data}
This study uses datasets from the MOOT repository (Multi-Objective Optimization Testing) \cite{ref29}, which compiles a diverse range of software engineering optimization problems from published research\footnote{https://github.com/timm/moot/tree/master/optimize}. The MOOT datasets are widely used in SE optimization studies as they cover tasks such as process tuning, configuration optimization, hyperparameter optimization (HPO), and management decision-making.

In total, we used 60 datasets categorized under five main types in MOOT:

\begin{itemize}[noitemsep, topsep=0pt]
    \item \textbf{Binary Configs.} This category contains datasets modeling binary configuration spaces, where each independent variable represents an on/off or enabled/disabled choice. These datasets are typically used to evaluate optimization techniques for highly configurable systems such as software product lines, where the configuration space can reach millions of possible combinations. Each dataset captures performance metrics such as runtime, memory usage, or defect counts for different feature combinations. 
    
    \item \textbf{Config.} Configuration datasets focus on system parameter tuning tasks. Examples include database configurations such as MySQL or Apache server settings. Each row represents a unique configuration with measured objectives like throughput and latency. For instance, tuning buffer sizes, thread pools, or cache settings to optimize database performance is a typical problem modeled in these datasets. These tasks are practically important, as default configurations are rarely optimal and manual tuning is often infeasible due to the size of the configuration space. 
    
    \item \textbf{HPO (Hyperparameter Optimization).} The HPO category includes datasets representing hyperparameter tuning tasks for machine learning algorithms. For example, optimizing the kernel type and regularization parameter for SVM, the number of estimators and maximum depth for Random Forest, or learning rate and boosting type for XGBoost. Each dataset captures the performance outcomes (e.g., accuracy, F1 score) for various hyperparameter settings, enabling comparative evaluation of optimization approaches for model tuning.
    
    \item \textbf{Process.} Process datasets capture software engineering process optimization problems, including task scheduling, effort estimation configurations, and process parameter tuning for builds or deployments. Objectives in these datasets often include minimizing total project duration, reducing cost, or minimizing defects introduced during processes. Such problems are critical for software project management and process improvement studies.

    \item \textbf{Misc.} The miscellaneous category consists of diverse datasets that do not fit other categories and are not related to software engineering. These datasets increase the diversity of optimization problems tested in this study. 
\end{itemize}

Across all categories, MOOT datasets are structured as tabular data with $x$ inputs (independent variables) and $y$ goals (dependent objectives). Numeric columns begin with uppercase letters, while symbolic columns begin with lowercase letters. Goal columns are annotated with “+” or “-” to indicate whether they are to be maximized or minimized, respectively. Dataset sizes vary from a few hundred to over 80,000 rows, with up to 88 independent variables and multiple objectives per dataset.

During experiments, rows are randomly shuffled, and $y$ values are hidden from the optimizer until queried, simulating realistic labeling cost constraints. Each optimizer request for a label incurs a unit cost, emphasizing the practical need for minimal-label optimization approaches. This diverse dataset collection provides a comprehensive evaluation ground to assess EZR’s ability to perform feature selection, optimization, and explanation effectively across a wide range of SE tasks. Table~\ref {tab:datasets-summary} summarizes the datasets used in this study. The table reports each dataset’s number of inputs and objectives (x/y), primary optimization objective, and file name.

\section{Results}
\subsection{How EZR is good at finding best settings?}
\subsubsection{Regression-Based Baselines}
\color{black}
To better understand the performance of EZR in selecting best configurations, we present our results across three separate tables, each grouping datasets by their input feature dimensionality. This stratification helps isolate how EZR performs under different levels of complexity.

\begin{itemize}[noitemsep, topsep=0pt]
    \item The first table contains results for datasets with \textbf{6 or fewer features}, referred to as \textit{light} datasets.
    \item The second table presents results for datasets with \textbf{between 7 and 19 features}, labeled as \textit{medium} datasets.
    \item The third table includes datasets with \textbf{20 or more features}, categorized as \textit{heavy} datasets.
\end{itemize}

Each column in the tables corresponds to a different regression model used in the optimization process. For each experimental run, the dataset is randomly split into 80\% training and 20\% test data. Models are trained on the training set and are tasked to identify the configuration in the test set that yields the lowest \textit{d2h} score. This process is repeated 20 times for each dataset, with a different random seed used in each repetition to account for sampling variability.

It is important to note that all models \textbf{except EZR} use labeled training data for learning. In contrast, \textbf{EZR operates under a label-constrained setting}, using only:
\begin{itemize}[noitemsep, topsep=0pt]
    \item \textbf{60} labeled configurations for \textit{light} and \textit{medium} datasets (labeled in two stages, first it selects 50 configurations from the pool of unlabeled settings, using active learning, and then it checks all the unlabeled data and labels the 10 best settings to find the actual best among them), and
    \item \textbf{10\% + 10}, 10\% of the training set for \textit{heavy} datasets at the first stage and 10 more samples at the next stage.
\end{itemize}

The choice of 50 sampling budget for light and medium datasets is grounded in established theory from optimization literature. Following the geometric process framework\cite{ref18} for near-optimal solution discovery, if solutions are randomly distributed and values within $\varepsilon$ are indistinguishable, the probability of drawing an $\varepsilon$-optimal solution follows a geometric distribution. The number of samples needed to achieve confidence $C$ of finding an $\varepsilon$-optimal solution is:
 \begin{equation}
1 - (1 - \varepsilon)^{n} \geq C
\label{eq:confidence-prob}
\end{equation}
which when solved for $n$ rearranges to
\begin{equation}
n(c, \varepsilon) = \frac{\log(1 - C)}{\log(1 - \varepsilon)}
\label{eq:confidence-prob2}
\end{equation}
  Cohen's rule~\cite{ref60} stats that results are statistically distinguishable if they differ by more than a ``small effect''.
 Here,  ``small'' and ``medium'' are  defined as
  ratios of a standard deviation; i.e.
  $0.2\varepsilon$ is ''small''; $0.5\varepsilon$ is ''medium''; and   $0.35\varepsilon$ is the border between small and medium.
  From an engineering perspective, Gaussian span the range $\pm 3\varepsilon$ so to 95\% confident of finding solutions indistinguishable from the best, this yields:
 \begin{equation}
n(C=0.95, \varepsilon=.35/6) \approx 50
\label{eq:samples-required}
\end{equation}
Similar theoretical reasoning has been employed by Lustosa et al.~\cite{ref17} in their partial-ordering explanations, drawing from Hamlet’s Probable Correctness Theory~\cite{ref18} and Hoare’s Quickselect pivot-and-partition strategy~\cite{ref19}, both of which support the use of small but statistically meaningful sampling budgets in optimization and explanation tasks. In addition to these theories, to validate the theoretical budget in practice, we also conducted experiments using a range of smaller and larger budget values, and the results of this sensitivity analysis are presented in Section~6.1.

Table \ref{tab:light-results-1}, \ref{tab:medium-results-1}, and \ref{tab:heavy-results-1} compare the performances of EZR with regression models for all datasets. The first row in each table lists the regression models used in our experiments. The column labeled \textit{AsIs} represents a random selection baseline, which simply chooses a configuration at random. The \textit{x} and \textit{y} columns indicate the number of input features and optimization objectives in each dataset, respectively. The \textit{x*} column shows the number of features presents in the EZR tree branch that leads to the best leaf. In other words, in order to find the best setting with EZR only that many features are actually used. The \textit{rows} column shows the total number of configurations in each dataset. The numbers reported in each table cell represent the \textbf{median accuracy} over 20 repeated runs, where the accuracy is measured as the win(\textit{d2h}) score of the best configuration identified in each run. Higher values indicate that the method consistently identified configurations closer to the ideal multi-objective trade-off. Green cells indicate the best performance achieved by any of the state-of-the-art regression models. Orange cells highlight cases where EZR finds a configuration whose performance is at least 90\% as good as the best model.

\begin{table}[t]
\scriptsize
\setlength{\tabcolsep}{3pt}
\begin{tabularx}{\columnwidth}{@{}ccccccY|cccccp{0.8cm}@{}}
\toprule
\textbf{AsIs} & \textbf{LR} & \textbf{RF} & \textbf{SVR} & \textbf{ANN} & \textbf{LGBM} & \textbf{EZR} & \textbf{data} & \textbf{x*} & \textbf{x} & \textbf{y} & \textbf{budget} & \textbf{rows} \\
\midrule
20            & \cellcolor[HTML]{B7E1CD}93  & \cellcolor[HTML]{B7E1CD}93  & \cellcolor[HTML]{B7E1CD}93  & 91                          & \cellcolor[HTML]{B7E1CD}93  & \cellcolor[HTML]{F9CB9C}92  & SS-A          & 3           & 3          & 2          & 60              & 1344          \\
-17           & \cellcolor[HTML]{B7E1CD}84  & \cellcolor[HTML]{B7E1CD}84  & \cellcolor[HTML]{B7E1CD}84  & \cellcolor[HTML]{B7E1CD}84  & \cellcolor[HTML]{B7E1CD}84  & \cellcolor[HTML]{B7E1CD}84  & SS-B          & 2           & 3          & 2          & 60              & 207           \\
25            & 80                          & \cellcolor[HTML]{B7E1CD}87  & 83                          & 79                          & \cellcolor[HTML]{B7E1CD}87  & \cellcolor[HTML]{F9CB9C}80  & SS-C          & 2           & 3          & 2          & 60              & 1513          \\
17            & 73                          & \cellcolor[HTML]{B7E1CD}85  & 84                          & 54                          & \cellcolor[HTML]{B7E1CD}85  & 72                          & SS-D          & 3           & 3          & 2          & 60              & 197           \\
25            & 82                          & \cellcolor[HTML]{B7E1CD}97  & \cellcolor[HTML]{B7E1CD}97  & 83                          & \cellcolor[HTML]{B7E1CD}97  & \cellcolor[HTML]{F9CB9C}91  & SS-E          & 3           & 3          & 2          & 60              & 757           \\
33            & 76                          & \cellcolor[HTML]{B7E1CD}91  & \cellcolor[HTML]{B7E1CD}91  & 73                          & \cellcolor[HTML]{B7E1CD}91  & \cellcolor[HTML]{F9CB9C}83  & SS-F          & 3           & 3          & 2          & 60              & 197           \\
23            & 76                          & \cellcolor[HTML]{B7E1CD}92  & \cellcolor[HTML]{B7E1CD}92  & 60                          & \cellcolor[HTML]{B7E1CD}92  & 81                          & SS-G          & 3           & 3          & 2          & 60              & 197           \\
-15           & 95                          & \cellcolor[HTML]{B7E1CD}99  & \cellcolor[HTML]{B7E1CD}99  & \cellcolor[HTML]{B7E1CD}99  & \cellcolor[HTML]{B7E1CD}99  & \cellcolor[HTML]{B7E1CD}99  & wc-1          & 2           & 3          & 1          & 60              & 197           \\
1             & 95                          & \cellcolor[HTML]{B7E1CD}99  & \cellcolor[HTML]{B7E1CD}99  & \cellcolor[HTML]{B7E1CD}99  & \cellcolor[HTML]{B7E1CD}99  & \cellcolor[HTML]{B7E1CD}99  & wc-2          & 2           & 3          & 1          & 60              & 197           \\
-2            & 95                          & \cellcolor[HTML]{B7E1CD}99  & \cellcolor[HTML]{B7E1CD}99  & 98                          & \cellcolor[HTML]{B7E1CD}99  & \cellcolor[HTML]{B7E1CD}99  & wc-3          & 2           & 3          & 1          & 60              & 197           \\
6             & \cellcolor[HTML]{B7E1CD}100 & \cellcolor[HTML]{B7E1CD}100 & \cellcolor[HTML]{B7E1CD}100 & \cellcolor[HTML]{B7E1CD}100 & \cellcolor[HTML]{B7E1CD}100 & \cellcolor[HTML]{B7E1CD}100 & SS-H          & 3           & 4          & 2          & 60              & 260           \\
18            & \cellcolor[HTML]{B7E1CD}98  & \cellcolor[HTML]{B7E1CD}98  & 42                          & \cellcolor[HTML]{B7E1CD}98  & \cellcolor[HTML]{B7E1CD}98  & \cellcolor[HTML]{B7E1CD}98  & SS-I          & 4           & 5          & 2          & 60              & 1081          \\
36            & \cellcolor[HTML]{B7E1CD}94  & \cellcolor[HTML]{B7E1CD}94  & \cellcolor[HTML]{B7E1CD}94  & 78                          & \cellcolor[HTML]{B7E1CD}94  & \cellcolor[HTML]{B7E1CD}94  & auto93        & 4           & 5          & 3          & 60              & 399           \\
-31           & 36                          & \cellcolor[HTML]{B7E1CD}64  & 49                          & 37                          & \cellcolor[HTML]{B7E1CD}64  & 42                          & HCI-hard      & 4           & 5          & 3          & 60              & 10001         \\
-22           & \cellcolor[HTML]{B7E1CD}100 & \cellcolor[HTML]{B7E1CD}100 & \cellcolor[HTML]{B7E1CD}100 & \cellcolor[HTML]{B7E1CD}100 & \cellcolor[HTML]{B7E1CD}100 & \cellcolor[HTML]{B7E1CD}100 & HCI-easy      & 3           & 5          & 3          & 60              & 10001         \\
-13           & 97                          & 96                          & 95                          & 96                          & \cellcolor[HTML]{B7E1CD}98  & \cellcolor[HTML]{F9CB9C}97  & SS-J          & 5           & 6          & 2          & 60              & 3841          \\
1             & 61                          & \cellcolor[HTML]{B7E1CD}98  & 94                          & 90                          & \cellcolor[HTML]{B7E1CD}98  & 81                          & SS-K          & 5           & 6          & 2          & 60              & 2881          \\
50            & 99                          & \cellcolor[HTML]{B7E1CD}100 & \cellcolor[HTML]{B7E1CD}100 & 94                          & \cellcolor[HTML]{B7E1CD}100 & \cellcolor[HTML]{F9CB9C}95  & SS-S          & 3           & 6          & 2          & 60              & 3841          \\
43            & \cellcolor[HTML]{B7E1CD}100 & \cellcolor[HTML]{B7E1CD}100 & 99                          & \cellcolor[HTML]{B7E1CD}100 & \cellcolor[HTML]{B7E1CD}100 & \cellcolor[HTML]{B7E1CD}100 & rs-1          & 4           & 6          & 1          & 60              & 3841          \\
92            & 99                          & \cellcolor[HTML]{B7E1CD}100 & 99                          & \cellcolor[HTML]{B7E1CD}100 & \cellcolor[HTML]{B7E1CD}100 & \cellcolor[HTML]{B7E1CD}100 & rs-2          & 3           & 6          & 1          & 60              & 3841          \\
-1            & 93                          & \cellcolor[HTML]{B7E1CD}95  & 94                          & 94                          & 94                          & \cellcolor[HTML]{F9CB9C}94  & sol-1         & 4           & 6          & 1          & 60              & 2867          \\
11            & \cellcolor[HTML]{B7E1CD}100 & \cellcolor[HTML]{B7E1CD}100 & \cellcolor[HTML]{B7E1CD}100 & \cellcolor[HTML]{B7E1CD}100 & \cellcolor[HTML]{B7E1CD}100 & \cellcolor[HTML]{B7E1CD}100 & wc-4          & 3           & 6          & 1          & 60              & 2881          \\
\bottomrule
\end{tabularx}
\caption{EZR Performance comparison for light datasets}
\label{tab:light-results-1}
\end{table}

As shown in Table \ref{tab:light-results-1}, which presents the results for the \textit{light} datasets (those with six or fewer features), EZR demonstrates competitive performance despite operating under strict label constraints. Specifically, EZR achieves at least 90\% of the best-known performance in 18 out of 22 datasets, accounting for 82\% of all cases. This is particularly noteworthy considering that EZR uses only 60 labeled examples per dataset, while all other regression models rely on full supervision using the entire training set. These results highlight EZR's effectiveness in identifying near-optimal configurations with minimal labeling effort, supporting its value in practical, label-scarce SE optimization tasks.

\begin{table}[t]
\scriptsize
\centering
\setlength{\tabcolsep}{3pt}
\begin{tabularx}{\columnwidth}{@{}ccccccY|cccccp{0.8cm}@{}}
\toprule
\rowcolor[HTML]{FFFFFF} 
\textbf{AsIs} & \textbf{LR} & \textbf{RF} & \textbf{SVR} & \textbf{ANN} & \textbf{LGBM} & \textbf{EZR} & \textbf{data} & \textbf{x*} & \textbf{x} & \textbf{y} & \textbf{budget} & \textbf{rows} \\
\midrule
\rowcolor[HTML]{FFFFFF} 
45            & \cellcolor[HTML]{B7E1CD}95  & \cellcolor[HTML]{B7E1CD}95  & \cellcolor[HTML]{B7E1CD}95  & \cellcolor[HTML]{B7E1CD}95  & \cellcolor[HTML]{B7E1CD}95  & \cellcolor[HTML]{B7E1CD}95  & Apache-AM     & 6           & 9          & 1          & 60              & 193           \\
-14           & \cellcolor[HTML]{B7E1CD}93  & 88                          & 87                          & 75                          & 88                          & 79                          & pom3a         & 4           & 9          & 3          & 60              & 20001         \\
-7            & \cellcolor[HTML]{B7E1CD}88  & 87                          & 87                          & 79                          & 87                          & 65                          & pom3b         & 4           & 9          & 3          & 60              & 20001         \\
-3            & \cellcolor[HTML]{B7E1CD}79  & 69                          & 78                          & 50                          & 68                          & 53                          & pom3c         & 5           & 9          & 3          & 60              & 20001         \\
6             & \cellcolor[HTML]{B7E1CD}89  & 87                          & 85                          & \cellcolor[HTML]{B7E1CD}89  & 85                          & 75                          & pom3d         & 5           & 9          & 3          & 60              & 501           \\
-10           & 55                          & 76                          & 76                          & 73                          & \cellcolor[HTML]{B7E1CD}77  & 54                          & Wine          & 5           & 10         & 2          & 60              & 1600          \\
-17           & \cellcolor[HTML]{B7E1CD}100 & 99                          & \cellcolor[HTML]{B7E1CD}100 & 98                          & \cellcolor[HTML]{B7E1CD}100 & \cellcolor[HTML]{F9CB9C}98  & SS-L          & 5           & 11         & 2          & 60              & 1024          \\
13            & 66                          & 96                          & 84                          & 96                          & \cellcolor[HTML]{B7E1CD}97  & \cellcolor[HTML]{F9CB9C}91  & SS-O          & 6           & 11         & 2          & 60              & 973           \\
-17           & \cellcolor[HTML]{B7E1CD}100 & 99                          & \cellcolor[HTML]{B7E1CD}100 & \cellcolor[HTML]{B7E1CD}100 & \cellcolor[HTML]{B7E1CD}100 & \cellcolor[HTML]{F9CB9C}99  & SS-P          & 7           & 11         & 2          & 60              & 1024          \\
10            & \cellcolor[HTML]{B7E1CD}98  & 95                          & \cellcolor[HTML]{B7E1CD}98  & \cellcolor[HTML]{B7E1CD}98  & \cellcolor[HTML]{B7E1CD}98  & 66                          & SS-X          & 4           & 11         & 2          & 60              & 86059         \\
62            & 87                          & \cellcolor[HTML]{B7E1CD}100 & 96                          & 88                          & \cellcolor[HTML]{B7E1CD}100 & \cellcolor[HTML]{F9CB9C}96  & SS-T          & 6           & 12         & 2          & 60              & 5185          \\
-30           & 63                          & \cellcolor[HTML]{B7E1CD}98  & 82                          & 97                          & \cellcolor[HTML]{B7E1CD}98  & \cellcolor[HTML]{B7E1CD}98  & SS-Q          & 6           & 13         & 3          & 60              & 2737          \\
51            & \cellcolor[HTML]{B7E1CD}100 & \cellcolor[HTML]{B7E1CD}100 & \cellcolor[HTML]{B7E1CD}100 & \cellcolor[HTML]{B7E1CD}100 & \cellcolor[HTML]{B7E1CD}100 & \cellcolor[HTML]{B7E1CD}100 & HSMGP         & 7           & 14         & 1          & 60              & 3457          \\
-28           & 81                          & \cellcolor[HTML]{B7E1CD}87  & \cellcolor[HTML]{B7E1CD}87  & 82                          & \cellcolor[HTML]{B7E1CD}87  & \cellcolor[HTML]{F9CB9C}80  & SS-R          & 7           & 14         & 2          & 60              & 3009          \\
12            & \cellcolor[HTML]{B7E1CD}97  & \cellcolor[HTML]{B7E1CD}97  & \cellcolor[HTML]{B7E1CD}97  & 96                          & \cellcolor[HTML]{B7E1CD}97  & \cellcolor[HTML]{B7E1CD}97  & SS-V          & 6           & 16         & 2          & 60              & 6841          \\

-5   & 54  & 70  & 66  & 54       & \cellcolor[HTML]{B7E1CD}99  & 53  & SS-W          & 7           & 16         & 2          & 60              & 65537          \\

15            & \cellcolor[HTML]{B7E1CD}100 & \cellcolor[HTML]{B7E1CD}100 & \cellcolor[HTML]{B7E1CD}100 & \cellcolor[HTML]{B7E1CD}100 & \cellcolor[HTML]{B7E1CD}100 & \cellcolor[HTML]{B7E1CD}100 & X264-AM       & 7           & 16         & 1          & 60              & 1153          \\
66            & \cellcolor[HTML]{B7E1CD}99  & \cellcolor[HTML]{B7E1CD}99  & \cellcolor[HTML]{B7E1CD}99  & \cellcolor[HTML]{B7E1CD}99  & \cellcolor[HTML]{B7E1CD}99  & \cellcolor[HTML]{B7E1CD}99  & SS-M          & 7           & 17         & 3          & 60              & 865           \\
9             & -5                          & \cellcolor[HTML]{B7E1CD}81  & 50                          & 58                          & \cellcolor[HTML]{B7E1CD}81  & 46                          & SS-N          & 6           & 17         & 2          & 60              & 53663        \\
\bottomrule
\end{tabularx}
\caption{EZR Performance comparison for medium datasets}
\label{tab:medium-results-1}
\end{table}

As shown in Table~\ref{tab:medium-results-1}, which reports the results for the \textit{medium} datasets (those with 7 to 19 input features), EZR continues to exhibit strong performance under label constraints. In this setting, EZR achieves at least 90\% of the best-known performance in 11 out of 19 datasets, accounting for 58\% of the cases. As with the \textit{light} datasets, EZR operates with only 60 labeled examples per dataset, while the other regression models are fully supervised using the entire training set.

\begin{table}[t]
\scriptsize
\centering
\setlength{\tabcolsep}{3pt}
\begin{tabularx}{\columnwidth}{@{}ccccccY|cccccp{0.8cm}@{}}
\toprule
\rowcolor[HTML]{FFFFFF} 
\textbf{AsIs} & \textbf{LR} & \textbf{RF} & \textbf{SVR} & \textbf{ANN} & \textbf{LGBM} & \textbf{EZR} 
  & \textbf{data} & \textbf{x*} & \textbf{x} & \textbf{y} & \textbf{budget} & \textbf{rows} \\ \hline
\rowcolor[HTML]{FFFFFF} 
-20           & \cellcolor[HTML]{B7E1CD}75  & 62                          & 57                          & 63                         & 56                          & 60                         & coc1000       & 5           & 20         & 5          & 60              & 1001          \\

60            & 97                          & \cellcolor[HTML]{B7E1CD}100 & 97                          & 97                         & \cellcolor[HTML]{B7E1CD}100 & \cellcolor[HTML]{F9CB9C}99 & SS-U          & 8           & 21         & 2          & 378             & 4609          \\
20            & 94                          & \cellcolor[HTML]{B7E1CD}96  & 87                          & 91                         & \cellcolor[HTML]{B7E1CD}96  & \cellcolor[HTML]{F9CB9C}93 & xomo-flight   & 7           & 27         & 4          & 810             & 10001         \\
13            & 96                          & \cellcolor[HTML]{B7E1CD}97  & 86                          & 86                         & \cellcolor[HTML]{B7E1CD}97  & \cellcolor[HTML]{F9CB9C}90 & xomo-ground   & 7           & 27         & 4          & 810             & 10001         \\
-6            & 92                          & \cellcolor[HTML]{B7E1CD}95  & 93                          & 93                         & \cellcolor[HTML]{B7E1CD}95  & \cellcolor[HTML]{F9CB9C}93 & xomo-osp      & 7           & 27         & 4          & 810             & 10001         \\
10            & \cellcolor[HTML]{B7E1CD}92  & 90                          & 82                          & 85                         & 91                          & \cellcolor[HTML]{F9CB9C}83 & xomo-osp2     & 8           & 27         & 4          & 810             & 10001         \\
-1            & 69                          & 74                          & 75                          & 71                         & \cellcolor[HTML]{B7E1CD}85  & 66                         & SQL-AM        & 11          & 39         & 1          & 382             & 4654          \\
-5            & 93                          & \cellcolor[HTML]{B7E1CD}98  & \cellcolor[HTML]{B7E1CD}98  & 84                         & \cellcolor[HTML]{B7E1CD}98  & \cellcolor[HTML]{B7E1CD}98 & billing10k    & 12          & 88         & 3          & 810             & 10001         \\
-4            & \cellcolor[HTML]{B7E1CD}97  & \cellcolor[HTML]{B7E1CD}97  & \cellcolor[HTML]{B7E1CD}97  & \cellcolor[HTML]{B7E1CD}97 & \cellcolor[HTML]{B7E1CD}97  & \cellcolor[HTML]{B7E1CD}97 & Scrum100k     & 15          & 124        & 3          & 8010            & 100001        \\
-19           & \cellcolor[HTML]{B7E1CD}95  & \cellcolor[HTML]{B7E1CD}95  & \cellcolor[HTML]{B7E1CD}95  & \cellcolor[HTML]{B7E1CD}95 & \cellcolor[HTML]{B7E1CD}95  & \cellcolor[HTML]{B7E1CD}95 & Scrum10k      & 12          & 124        & 3          & 810             & 10001         \\
10            & \cellcolor[HTML]{B7E1CD}95  & \cellcolor[HTML]{B7E1CD}95  & \cellcolor[HTML]{B7E1CD}95  & \cellcolor[HTML]{B7E1CD}95 & \cellcolor[HTML]{B7E1CD}95  & \cellcolor[HTML]{F9CB9C}86 & Scrum1k       & 8           & 124        & 3          & 90              & 1001          \\
9             & \cellcolor[HTML]{B7E1CD}99  & \cellcolor[HTML]{B7E1CD}99  & \cellcolor[HTML]{B7E1CD}99  & \cellcolor[HTML]{B7E1CD}99 & \cellcolor[HTML]{B7E1CD}99  & \cellcolor[HTML]{B7E1CD}99 & FFM-125       & 8           & 128        & 3          & 810             & 10001         \\
24            & \cellcolor[HTML]{B7E1CD}100 & \cellcolor[HTML]{B7E1CD}100 & \cellcolor[HTML]{B7E1CD}100 & 93                         & \cellcolor[HTML]{B7E1CD}100 & \cellcolor[HTML]{F9CB9C}97 & FFM-250       & 12          & 256        & 3          & 810             & 10001         \\
11            & \cellcolor[HTML]{B7E1CD}95  & \cellcolor[HTML]{B7E1CD}95  & \cellcolor[HTML]{B7E1CD}95  & 93                         & \cellcolor[HTML]{B7E1CD}95  & \cellcolor[HTML]{B7E1CD}95 & FFM-500       & 13          & 510        & 3          & 810             & 10001         \\
11            & \cellcolor[HTML]{B7E1CD}93  & \cellcolor[HTML]{B7E1CD}93  & \cellcolor[HTML]{B7E1CD}93  & \cellcolor[HTML]{B7E1CD}93 & \cellcolor[HTML]{B7E1CD}93  & \cellcolor[HTML]{B7E1CD}93 & FM-500-1      & 11          & 511        & 3          & 810             & 10001         \\
22            & \cellcolor[HTML]{B7E1CD}97  & \cellcolor[HTML]{B7E1CD}97  & \cellcolor[HTML]{B7E1CD}97  & \cellcolor[HTML]{B7E1CD}97 & \cellcolor[HTML]{B7E1CD}97  & \cellcolor[HTML]{B7E1CD}97 & FM-500-2      & 11          & 511        & 3          & 810             & 10001         \\
9             & \cellcolor[HTML]{B7E1CD}95  & \cellcolor[HTML]{B7E1CD}95  & \cellcolor[HTML]{B7E1CD}95  & \cellcolor[HTML]{B7E1CD}95 & \cellcolor[HTML]{B7E1CD}95  & \cellcolor[HTML]{B7E1CD}95 & FM-500-3      & 9           & 513        & 3          & 810             & 10001         \\
0             & \cellcolor[HTML]{B7E1CD}97  & \cellcolor[HTML]{B7E1CD}97  & \cellcolor[HTML]{B7E1CD}97  & 96                         & \cellcolor[HTML]{B7E1CD}97  & \cellcolor[HTML]{F9CB9C}96 & FM-500-4      & 14          & 517        & 3          & 810             & 10001         \\
17            & 86                          & \cellcolor[HTML]{B7E1CD}94  & \cellcolor[HTML]{B7E1CD}94  & 86                         & \cellcolor[HTML]{B7E1CD}94  & \cellcolor[HTML]{B7E1CD}94 & FFM-1000      & 11          & 1044       & 3          & 810             & 10001        \\
\bottomrule
\end{tabularx}
\caption{EZR Performance comparison for heavy datasets}
\label{tab:heavy-results-1}
\end{table}

As shown in Table~\ref{tab:heavy-results-1}, which presents the results for the \textit{heavy} datasets (those with more than 20 input features), EZR achieves its strongest performance under label constraints. In this group, EZR reaches at least 90\% of the best-known performance in 17 out of 19 datasets, amounting to 89\% of all cases—its highest success rate across all dataset categories. Notably, for these larger datasets, EZR uses only 10 + 10\% of the available training labels.\footnote{Except for coc1000 and scrum1k, which have a labeling budget of 60 since they have fewer rows.} This high efficiency is particularly significant because these heavy datasets are more representative of real-world scenarios, where the configuration space is large, the number of tunable parameters is high, and labeling costs are substantial. EZR’s ability to perform well in such settings highlights its practicality for real-world software engineering optimization tasks.

To be more specific, Table~\ref{tab:agg_results} aggregates EZR's performance across all 60 datasets used in this study. The first column shows the number of datasets and its corresponding proportion relative to the full datasets where EZR reaches a given performance threshold. The second column indicates the minimum percentage efficiency (relative to the best-performing answer) that EZR achieved in those cases. Notably, in 55 out of all 60 datasets (92\%), EZR was able to find a configuration yielding at least 70\% of the performance of the best-known method. Even under stricter criteria, EZR achieved at least 90\% of the best-known performance in 46 datasets (77\%), demonstrating strong label efficiency and practical competitiveness.

\begin{table}
\centering
\scriptsize
\begin{tabular}{cc}
\toprule
\textbf{\# Datasets} &   \textbf{Relative Efficiency} \\
\midrule
55 / 60 =  92\% & $\geq$ 70\% \\
52 / 60 =  87\% & $\geq$ 80\% \\
46 / 60 =  77\% & $\geq$ 90\% \\
\bottomrule
\end{tabular}
\caption{EZR Performance across all 60 datasets.}
\label{tab:agg_results}
\end{table}
 
To provide statistical rigor, we incorporated the Scott-Knott (SK) test, a hierarchical clustering procedure designed for multiple treatment comparisons in software engineering research\cite{ref30}. SK recursively partitions treatments into statistically distinct ranks by maximizing between-group differences while minimizing within-group variance. Unlike pairwise tests that suffer from multiple comparison problems, SK produces non-overlapping groups where treatments within the same rank are statistically indistinguishable.
We conducted two SK analyses with different confidence thresholds. As shown in Table~\ref{tab:sk_results}, using a stringent threshold(epsilon), EZR does not always achieve the top statistical rank occupied by fully supervised methods, as expected given EZR's severe label constraints. However, with a relaxed threshold(10\% of the best achieved performance), EZR consistently clustered within the same statistical rank as state-of-the-art methods in the \textbf{majority} of datasets. This demonstrates that while EZR may not always match the absolute best performer under strictest criteria, the performance gap is statistically negligible under wider thresholds, substantiating our claim that EZR "consistently achieves near-optimal performance."
\begin{table}

\centering
\scriptsize
\begin{tabular}{cccc}
\toprule
 & \textbf{Light(22)} &   \textbf{Medium(19)} & \textbf{Heavy(19)}\\
\midrule
\textbf{Tight Threshold} & 11/22  &  6/19  &  10/19 \\
\textbf{Wider Threshold} & 19/22  &  12/19  &  17/19 \\
\bottomrule
\end{tabular}
\caption{On how many datasets in each group does EZR perform statistically indifferent compared to the best achieved performance?}
\label{tab:sk_results}
\end{table}

\subsubsection{Optimization Baselines}
To directly evaluate EZR against established optimization algorithms, we compared EZR against two state-of-the-art optimizers, SMAC and DEHB, under two fixed evaluation budgets:
\begin{itemize}
    \item Labeling Budget = 50
    \item Labeling Budget = 200
\end{itemize}
All methods (EZR, SMAC, DEHB, and Random) were restricted to the same number of objective evaluations to ensure fairness.

\begin{figure}[t]
  \centering
  \begin{minipage}{0.49\linewidth}
    \centering
  \fbox{\includegraphics[width=\textwidth]{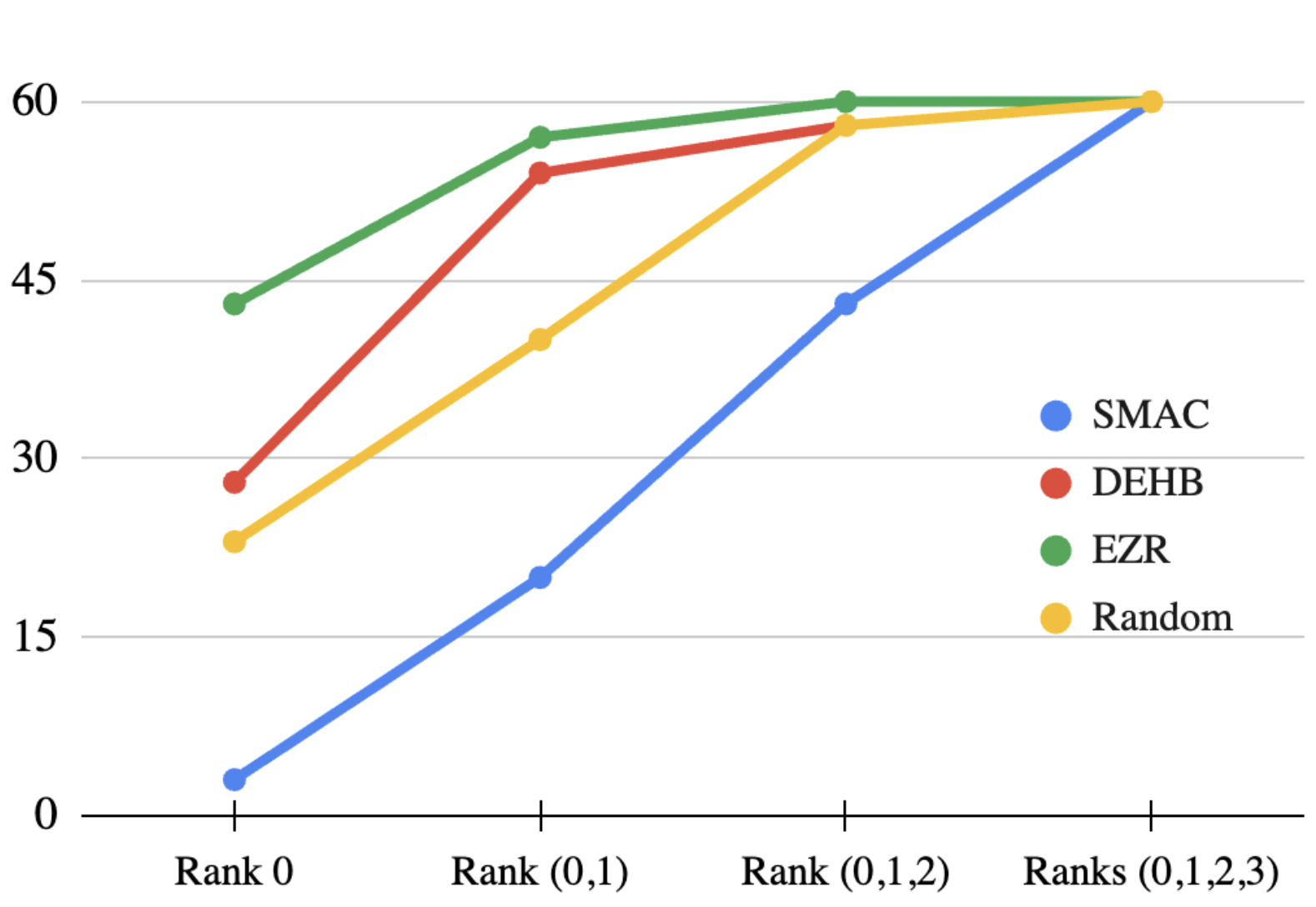}}
    \caption{Among 60 datasets, how many times each method has statistically better performance, using 50 labels? }
    \label{fig:budget-small}
  \end{minipage}\hfill
  \begin{minipage}{0.47\linewidth}
    \centering
  \fbox{\includegraphics[width=\textwidth]{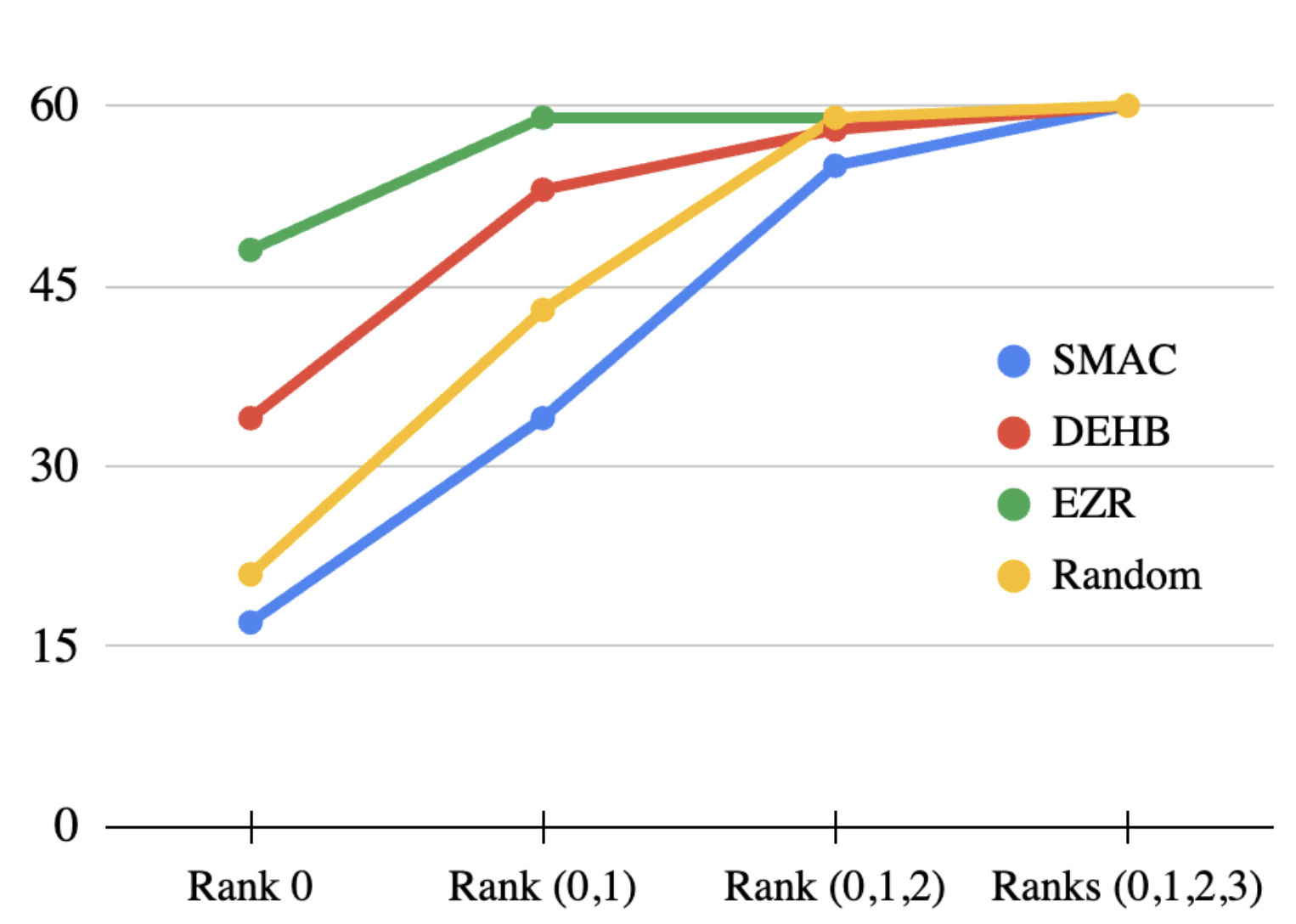}}
    \caption{Among 60 datasets, how many times each method has statistically better performance, using 200 labels? }
    \label{fig:budget-heavy}
  \end{minipage}
\end{figure}

Figures X and Y summarize the statistical rankings across the 60 datasets for budgets of 50 and 200 evaluations, respectively. The y-axis reports how many datasets each method was statistically indistinguishable from the top rank. Under the stricter budget of 50 evaluations, EZR achieves the top statistical rank in substantially more datasets than SMAC and DEHB. EZR is ranked in the best statistical group for nearly all datasets, while SMAC lags significantly under this tight budget. DEHB performs better than SMAC but remains below EZR in the number of top-ranked outcomes. With a larger budget of 200 evaluations, performance gaps narrow, as expected. All methods improve with additional evaluations. However, EZR continues to match or exceed the state-of-the-art methods across most datasets, remaining statistically competitive in nearly all cases.

These results show that EZR performs strongly under strict labeling budgets, consistently matching or outperforming SMAC and DEHB across the majority of datasets. Even with only 50 evaluations, EZR frequently achieves statistically superior or indistinguishable performance compared to state-of-the-art optimizers. Increasing the budget to 200 evaluations reduces performance differences across methods, but EZR remains competitive.

\color{black}

\color{black}
Looking back to RQ1, we say:
\begin{quote}
{\bf  Answering RQ1:  Effectiveness on optimization.} {\em The results demonstrate that EZR consistently achieves near-optimal performance across a wide range of datasets while using only a fraction of the labeling effort required by fully supervised models. This is a clear answer to RQ1; \textbf{EZR is an effective optimizer under label-scarce conditions, rivaling or exceeding state-of-the-art regression \color{black} and optimization \color{black} methods despite its minimal data requirements.} These findings reinforce the \textbf{Maximum Clarity Heuristic} introduced earlier—showing that, in complex software engineering optimization tasks, strong results can be achieved with less data when guided by clear and efficient heuristics.}
\end{quote}

\subsection{To what extent are EZR's results explainable?}
Understanding the outcomes of optimization in software engineering requires not only identifying optimal solutions but also interpreting \emph{why} those solutions are preferred. This calls for XAI techniques that can provide insight into both the global behavior of the model and its local decision processes. RQ2 and RQ3 investigate the extent to which EZR provides explainable outputs compared to established XAI methods. Explainability is inherently multi-faceted: an explanation should be understandable to humans while also faithfully reflecting the underlying predictive signals. To capture both dimensions, we evaluate EZR using two complementary approaches. First, we present a subjective comparison of EZR’s local and global explanations against widely used methods. Second, we conduct an objective evaluation, where we test whether the feature-importance rankings produced by each explanation method can guide effective feature selection and improve downstream optimization. Together, these analyses provide a holistic assessment of EZR’s explainability.

\subsubsection{Explanations in Practice: What Practitioners See} \label{subjectiveXAI}
Explainability is particularly critical in software engineering, where developers, project managers, and stakeholders must understand model behavior to justify decisions and ensure accountability. Without clear explanations, even high-performing models may face resistance in practice \cite{ref22}. Explanations can be understood at two levels, \textit{global explanations}, which reveal overall feature importance across the dataset, and \textit{local explanations}, which clarify why the model produced a specific outcome for an individual instance. In the following, we present both global and local outputs of EZR and compare them against established methods.

\begin{figure}[htbp]
  \centering
  \includegraphics[width=0.75\textwidth]{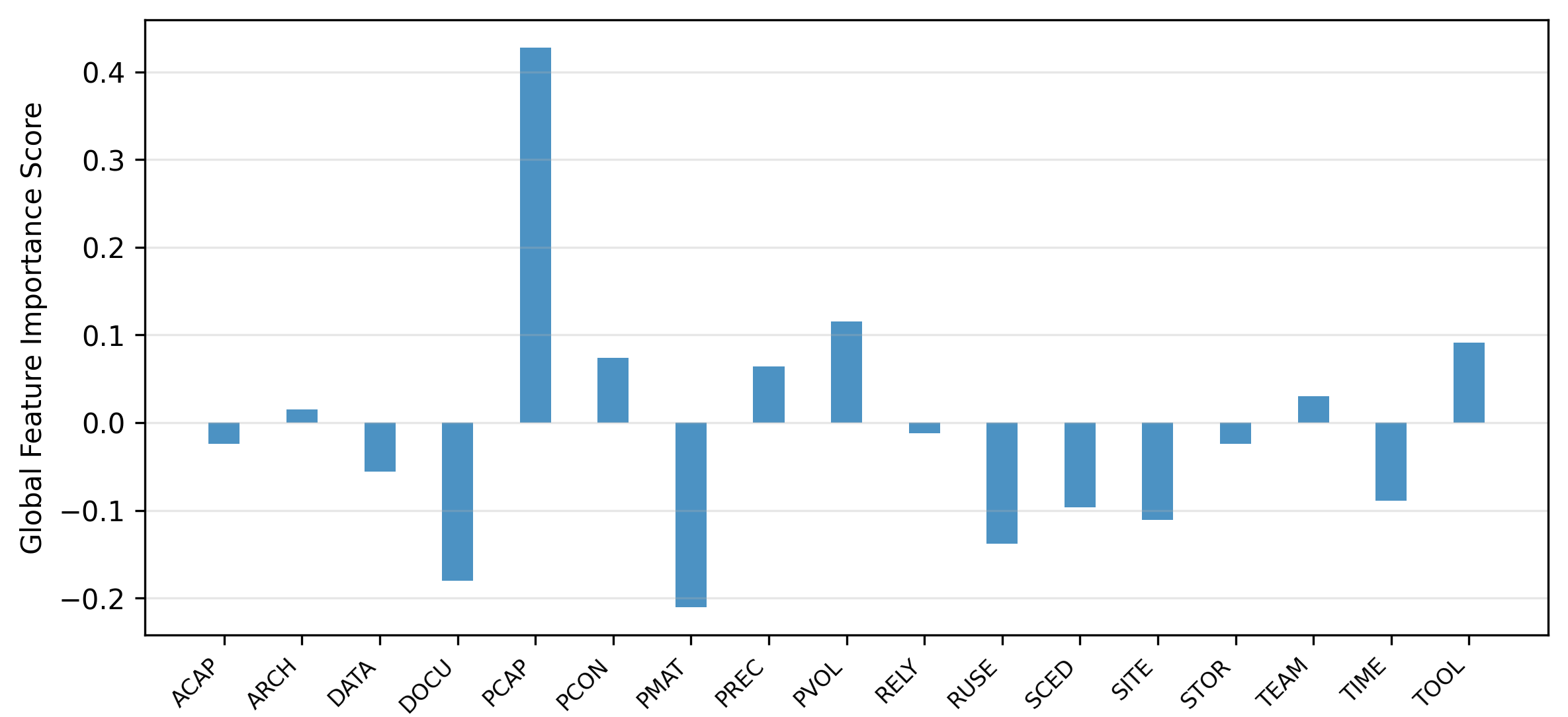}
  \caption{COC1000 Feature importance via permutation}
  \label{fig:perm-importance}
\end{figure}

\textbf{Global Explanation:}
For global explanations, we adopt the permutation importance framework, a widely used model-agnostic technique for quantifying feature contributions. Following Alcaide et al. \cite{ref24}, who demonstrated its utility for configuration-error detection, we use this approach to benchmark EZR. Figure \ref{fig:perm-importance} presents the resulting permutation importance values (y-axis). A large degradation in performance indicates that a feature is highly informative, while negative importance values suggest that a feature’s original distribution may add noise or reduce predictive accuracy.

\begin{figure}[htbp]
\scriptsize
\centering
\begin{lstlisting}
 win    n
---- ----
   6   32
  12   30  STOR <= 5
  13   28  |  TEAM <= 5
  20   25  |  |  TEAM <= 4
  22   23  |  |  |  PREC <= 5
  34    2  |  |  |  |  TEAM <= 1 ;
  21   21  |  |  |  |  TEAM >  1
  41    2  |  |  |  |  |  PVOL <= 2 ;
  18   19  |  |  |  |  |  PVOL >  2
  65    4  |  |  |  |  |  |  PCON <= 1
  @76    2  |  |  |  |  |  |  |  ACAP >  4 ;@
  55    2  |  |  |  |  |  |  |  ACAP <= 4 ;
   6   15  |  |  |  |  |  |  PCON >  1
  48    2  |  |  |  |  |  |  |  PREC <= 2 ;
  -1   13  |  |  |  |  |  |  |  PREC >  2
   8    2  |  |  |  |  |  |  |  |  RUSE <= 2 ;
  -2   11  |  |  |  |  |  |  |  |  RUSE >  2
   0    9  |  |  |  |  |  |  |  |  |  PVOL <= 4
  16    2  |  |  |  |  |  |  |  |  |  |  PCON >  4 ;
  -5    7  |  |  |  |  |  |  |  |  |  |  PCON <= 4
  17    2  |  |  |  |  |  |  |  |  |  |  |  RUSE >  5 ;
 -14    5  |  |  |  |  |  |  |  |  |  |  |  RUSE <= 5
  -9    2  |  |  |  |  |  |  |  |  |  |  |  | DOCU <= 2;
 -17    3  |  |  |  |  |  |  |  |  |  |  |  | DOCU >  2;
 -11    2  |  |  |  |  |  |  |  |  |  PVOL >  4 ;
  -3    2  |  |  |  PREC >  5 ;
 -45    3  |  |  TEAM >  4 ;
  -1    2  |  TEAM >  5 ;
 -79    2  STOR >  5 ;
\end{lstlisting}
\caption{ EZR  tree output showing impurity-based splits and performance metrics at each node.}
\label{fig:ezr-tree-output}
\end{figure}

As discussed earlier, to solve optimization problems, our method \textbf{ EZR } constructs an interpretable decision model using only a small number of labeled examples. Figure \ref{fig:ezr-tree-output} presents a Parzen-style decision tree constructed by EZR\footnote{Extra details are omitted for brevity. For the whole tree, refer to \href{https://github.com/amiiralii/Minimal-Data-Maximum-Clarity/tree/main/paper_materials}{https://github.com/amiiralii/Minimal-Data-Maximum-Clarity/tree/main/paper\_materials}}, where each row corresponds to a node that recursively splits the configuration space. Internal nodes divide the data based on feature thresholds (e.g., \textit{TEAM <= 5}), and leaf nodes are indicated by a semicolon (\textit{;}). Each node includes the number of configurations it contains (\textit{n}) and the \textit{win} score of the median performance in terms of \textit{d2h}. As an example, the most optimal leaf in this tree corresponds to a branch with only 2 configurations (out of all those intelligently chosen ones) and a win score of 76\%. To identify high-performing configurations in practice, we would pass any test instances through the tree and observe which branch they fall into. We then inspect the configurations in that branch and label the top-ranked ones to determine which yields the best actual \textit{d2h}. Also, based on the EZR tree and by using MDI (Equation \ref{eq:mdi}), we can extract feature importance scores. This use case will be discussed in section \textit{4.2.2}. 

\textbf{Local Explanation}
To illustrate local explanations of common baselines, we use a sample test instance from the COC1000 dataset (Table\ref{tab:instance}) and predict its \textit{d2h} value with a LightGBM model. For this instance, we generate explanations using three widely adopted techniques: \textbf{LIME}, \textbf{SHAP}, and \textbf{BreakDown}. In contrast, EZR provides local explanations natively. Its decision tree structure allows tracing the decision path for a given configuration, thereby revealing why the model identifies it as promising or not.

\begin{table}[htbp]
  \centering
  \scriptsize
  \begin{tabular}{@{}*{17}{c}@{}}
    \toprule
    ACAP & \cellcolor[HTML]{EFEFEF} ARCH & DATA & \cellcolor[HTML]{EFEFEF} DOCU &
    PCAP & \cellcolor[HTML]{EFEFEF} PCON & PMAT & \cellcolor[HTML]{EFEFEF} PREC & PVOL \\
    2    & \cellcolor[HTML]{EFEFEF} 5    & 3    & \cellcolor[HTML]{EFEFEF} 1    &
    3    & \cellcolor[HTML]{EFEFEF} 1    & 3    & \cellcolor[HTML]{EFEFEF} 5   &   3   \\
    \midrule
    \cellcolor[HTML]{EFEFEF} RELY & RUSE & \cellcolor[HTML]{EFEFEF} SCED &
    SITE & \cellcolor[HTML]{EFEFEF} STOR & TEAM & \cellcolor[HTML]{EFEFEF} TIME &
    TOOL \\
    \cellcolor[HTML]{EFEFEF} 1    & 2    & \cellcolor[HTML]{EFEFEF} 2    &
    4    & \cellcolor[HTML]{EFEFEF} 3    & 3    & \cellcolor[HTML]{EFEFEF} 5    &
    3    \\
    \bottomrule
  \end{tabular}
  \caption{Feature values for the selected test instance}
  \label{tab:instance}
\end{table}

Starting with LIME's results available in Figure~\ref{fig:Lime output}, the surrogate model attributes the prediction (0.50 vs. true 0.45) mainly to \textbf{$SCED \leq 2$} and \textbf{$ACAP \leq 2$}, both of which push the value upward. In contrast, \textbf{$2 < PMAT \leq 4$} exerts the strongest downward effect, indicating that moderate process maturity improves the configuration. Considering smaller contributions from other features, together these attributions explain the adjustment from the local baseline to the final prediction.

\begin{figure}[ht]
\centering
\begin{lstlisting}
True target = 0.45, Model prediction = 0.50
LIME Explain:
        SCED <= 2.00: +0.017 (↑)
        ACAP <= 2.00: +0.010 (↑)
 2.00 < PMAT <= 4.00: -0.006 (↓)
 4.00 < TIME <= 5.00: +0.006 (↑)
 4.00 < PREC <= 5.00: +0.003 (↑)
        DOCU <= 2.00: +0.003 (↑)
\end{lstlisting}
\caption{LIME explanation output}
\label{fig:Lime output}
\end{figure}

\begin{figure}[!t]
  \centering
  \includegraphics[width=0.7\textwidth]{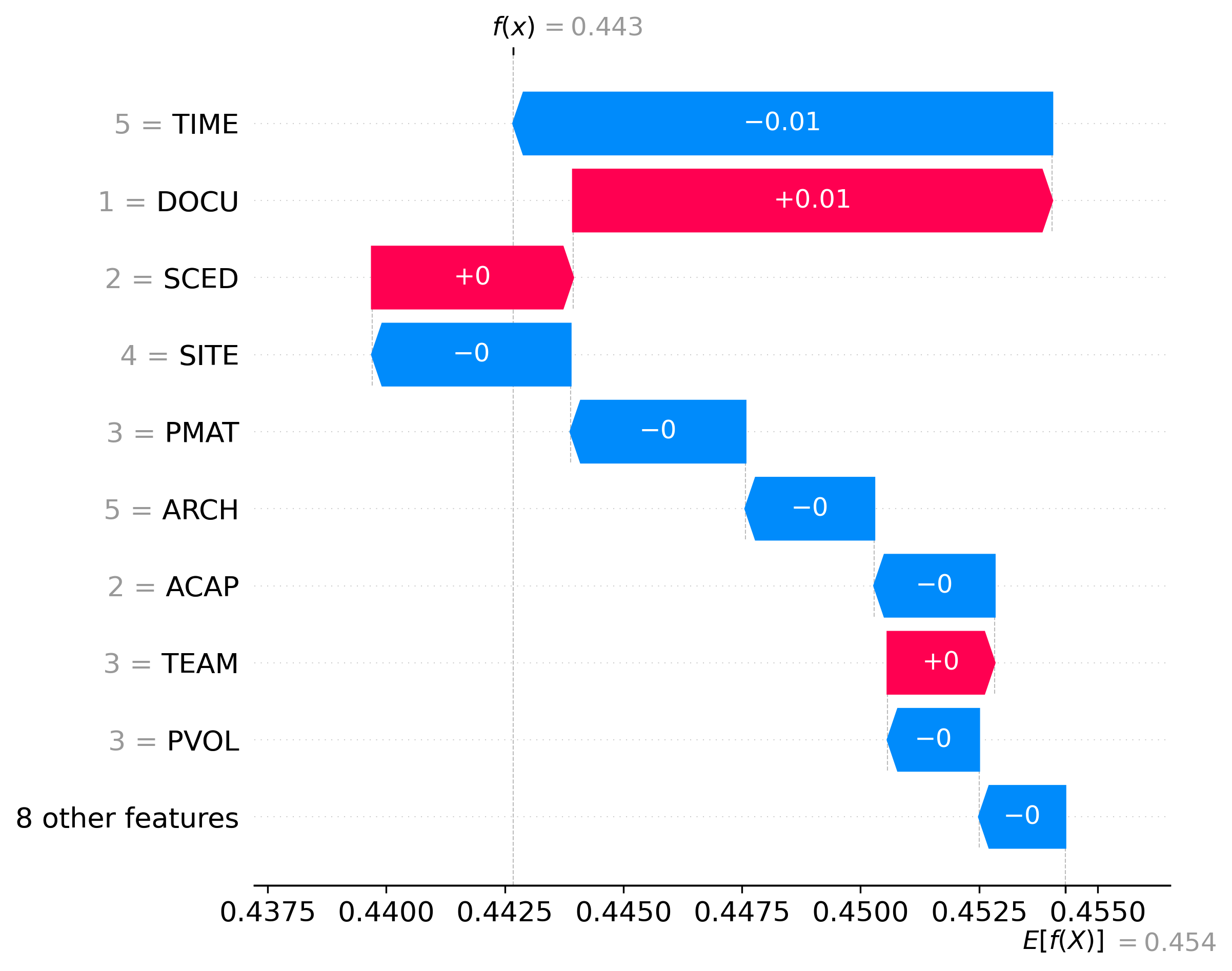}
  \caption{SHAP explanation output}
  \label{fig:shap output}
\end{figure}

 Moving on to SHAP, Figure \ref{fig:shap output} shows the output of SHAP's waterfall results. It begins at the baseline expectation $E[f(X)] = 0.454$ and then sees how each feature nudges the prediction down to $f(x) = 0.443$. The largest driver is TIME, which strongly reduces the \textit{d2h} value, while DOCU raises it slightly. Other factors such as PMAT, ARCH, and ACAP each provide small downward contributions, whereas TEAM nudges the value upward. Overall, SHAP highlights TIME and DOCU as the dominant influences for this instance.

\begin{figure}[b]
  \centering
  \includegraphics[width=\textwidth]{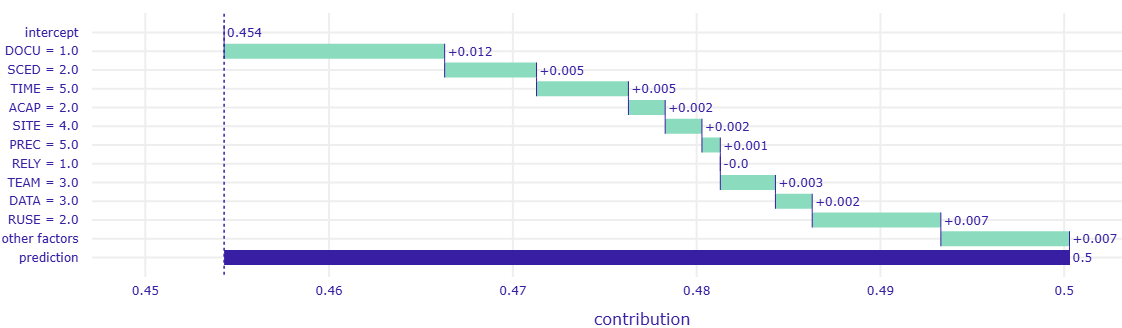}
  \caption{BreakDown explanation output}
  \label{fig:breakdown output}
\end{figure}
For BreakDown, Figure~\ref{fig:breakdown output}, the prediction builds sequentially from the baseline 0.454 to 0.500. The largest contributors are DOCU,  SCED, and TIME, with smaller additions from ACAP, SITE, and others. This stepwise view highlights both the dominant drivers and how their cumulative effects produce the final prediction.

For EZR’s local explanation, we trace the decision path from the root to the final leaf of the tree for the given instance (Figure~\ref{fig:ezr-tree-branch}). The objective is not to monitor incremental prediction changes but to identify the leaf that groups together configurations sharing similar characteristics with the instance. In this case, the path terminates in a leaf containing only two configurations, providing a cohort-based rationale: \textit{this configuration behaves like those cases}. Our implementation also exposes the instances within each leaf along with summary statistics (e.g., mean \textit{d2h}, variance, win percentage, Impurity), enabling direct inspection of how comparable cases performed in practice. It is important to note that this path involves only six features, while the full EZR tree uses nine drawn from seventeen. This reduction underscores EZR’s ability to highlight only the most informative attributes, which aligns with our \textbf{Maximum Clarity Heuristic} that complex tasks can often be explained with less data.

\begin{figure}[!t]
\centering
\begin{lstlisting}
 win    n 
---- ---- 
   6   32
  12   30  STOR <= 5
  13   28  |  TEAM <= 5
  20   25  |  |  TEAM <= 4
  22   23  |  |  |  PREC <= 5
  21   21  |  |  |  |  TEAM >  1
  18   19  |  |  |  |  |  PVOL >  2
  65    4  |  |  |  |  |  |  PCON <= 1
  55    2  |  |  |  |  |  |  |  ACAP <= 4 ;
\end{lstlisting}
\caption{ Highlighted decision path in the EZR tree}
\label{fig:ezr-tree-branch}
\end{figure}

Having presented the global and local outputs of EZR alongside established methods such as LIME, SHAP, and BreakDown, we now turn to a subjective comparison of their explanatory value. To structure this assessment, we draw on the explanation triggers and user-goal framework (Table \ref{tab:xai_triggers}), which captures the types of questions practitioners typically pose when interacting with AI systems. By examining how well each method addresses these triggers, we can evaluate their relative clarity, actionability, and alignment with stakeholder needs.

\begin{table}[htbp]
\scriptsize
\centering
\setlength{\tabcolsep}{6pt}
\begin{tabularx}{\textwidth}{
  >{\raggedright\arraybackslash}p{0.32\textwidth}
  >{\raggedright\arraybackslash}p{0.4\textwidth}
  >{\raggedright\arraybackslash}X
}
\textbf{Trigger} & \textbf{User/Learner's Goal} & \textbf{Rung in ladder of causation} \\
\midrule
\rowcolor[HTML]{EFEFEF} How do I use it? & Achieve the primary task goals & --- \\
How does it work? & Satisfaction from achieving global understanding  & Association \\
\rowcolor[HTML]{EFEFEF} What did it just do? & Satisfaction from understanding local decision & Intervention\\
What does it achieve? & Understand the system's functions and use & Intervention \\
\rowcolor[HTML]{EFEFEF} What will it do next? & Trust based on observability and predictability & Intervention\\
How much effort will this take? & Effectiveness in achieving primary goals & Intervention\\
\rowcolor[HTML]{EFEFEF} What do I do if it gets it wrong? & Avoid mistakes & Counterfactual\\
How do I avoid the failure modes? & Mitigate errors & Counterfactual \\
\rowcolor[HTML]{EFEFEF} What would it have done if \textit{x} were different? & Resolve curiosity through counterfactual understanding & Counterfactual\\
Why didn’t it do \textit{z}? & Understand the rationale behind a local decision & Counterfactual\\
\bottomrule
\end{tabularx}
\caption{Common explanation triggers and associated user goals in XAI systems \cite{ref21} with corresponding rung in the ladder of causation \cite{ref53}}
\label{tab:xai_triggers}
\end{table}

Building on prior work advocating the integration of causal inference into software engineering \cite{ref52}, we also evaluate \textbf{EZR} through Pearl’s Ladder of Causation \cite{ref53}, which distinguishes three levels of reasoning; \textbf{seeing} (associations), \textbf{doing} (interventions), and \textbf{imagining} (counterfactuals). Attribution methods largely remain on the first rung, providing associational insight by decomposing a prediction into additive contributions ($P(Y|T)$) but without modeling interventions or alternative realities. EZR, by contrast, spans all three rungs. At the associational level, its global tree structure and permutation importance summarize which features matter overall. At the interventional level, EZR natively supports “what-if” reasoning. Each split is a concrete threshold, so modifying a feature and retracing the path yields a new leaf (cohort) and updated outcome summaries, directly operationalizing the do-operator ($P(Y|do(T))$). Finally, at the counterfactual level, minimal edits to feature values identify the closest alternative leaf where the decision would differ, enabling users to ask not only “what if $x$ were different?” but also “why didn’t it do $z$?”. Viewed through the practitioner triggers in Table~\ref{tab:xai_triggers}, this causal mapping strengthens our subjective evaluation: EZR addresses global needs (“How does it work?”), local reasoning (“What did it just do?”), and counterfactual queries (“what-if/why”) in a unified and deterministic framework. In doing so, it delivers explanations that are interpretable, consistent, and operational.

To further illustrate EZR’s explanatory capability, we perform a walkthrough using the sample configuration from Table 7. We evaluate how well the cohort-based explanation addresses each practitioner question outlined in Table 8. 
\begin{enumerate}[noitemsep, topsep=0pt]
    \item \textbf{How do I use it?} 
    Feed the instance into the EZR tree (Figure 3) to trace its decision path. Looking at the win rate of the leaf the configuration ended up in, you can decide how good the instance is.
    \item \textbf{How does it work?} 
    EZR splits the configuration space using the most influential features first, those appearing near the root or repeatedly across branches. The MDI score quantifies this influence, offering a global view of which features consistently shape decisions. In practice, this provides a quick, interpretable summary of what matters most in the system.
    \item \textbf{What did it just do?}, \textbf{What does it achieve?}
    EZR evaluated the instance by following a path through key splits, for example, $TEAM \leq 4$ suggests a smaller, likely more cohesive team, or $PREC \leq 5$ implies moderate process flexibility. Based on these and other decisions(Figure \ref{fig:ezr-tree-branch}), the configuration was assigned to a cohort with a win score of 55, meaning it performs reasonably well but not among the top-tier options. In practice, this points to a viable but improvable setup for project planning.    
    \item \textbf{What will it do next?}
    Based on the current cohort’s win score of 55 (vs. a peak of 76 in the tree), EZR infers that this configuration is unlikely to be optimal. Rather than labeling or deploying it, the system recommends exploring other configurations with higher potential. Also, in order to end up in the best known leaf, EZR suggests increasing ACAP, corresponding to having more skilled analysts for the project.
    \item \textbf{How much effort will this take?}
    EZR requires minimal effort. It simply routes the new configuration through the tree in milliseconds. If the outcome is suboptimal, users can review the decision path, identify the feature that steered the configuration into a low-performing branch (e.g., ACAP < 4), and tweak that setting to redirect it toward a better-performing cohort. Thus, no retraining or full model inspection needed.
    \item \textbf{What do i do if it gets wrong?}, \textbf{How do i avoid the failure modes?}
    Practitioners can mitigate risks in three ways. First, remain cautious of dataset biases. If most high-performing samples share specific traits(for instance, favoring small teams if most high-performing examples come from solo projects), EZR’s tree may overemphasize them; inspecting second-best cohorts can reveal more balanced alternatives. Second, revisit the labeling budget. Although we adopted values recommended by prior work, some problems may require more labels for stability, while in other cases fewer may suffice. So adaptive adjustments are encouraged. Finally, when a configuration is assigned to a leaf but lies far from that leaf’s centroid (compared to its peers), the cohort rationale may be unreliable. In such cases, users should treat EZR’s recommendation as tentative and seek validation from nearby leaves or additional labeling.
    \item \textbf{What would it have done if $x$ were different?}
    EZR enables easy what-if reasoning. For our sample configuration, increasing ACAP (Analyst Capability) would redirect it from a lower performing branch (win = 55) to a higher one (win = 76). This suggests that boosting analyst expertise, perhaps through targeted hiring or internal training programs, could significantly improve project outcomes. Such actionable insight helps decision-makers connect model recommendations to real-world interventions.
    \item \textbf{Why didn't it do $z$?}
    Answering counterfactual questions is straightforward with EZR, thanks to its transparent decision tree. For example, “Why isn’t this configuration among the best?” Because its ACAP (Analyst Capability) is too low, pushing it into a lower-performing branch. On the other hand, a higher-scoring configuration satisfies key thresholds such as $STOR \leq 5$ indicating sufficient storage resources, $TEAM \leq 4$ reflecting a cohesive small team structure, $PREC \leq 5$ suggesting moderate flexibility in development processes (Precedentedness), $PCON \leq 1$, implying low risk of personnel change (Personnel Continuity), and $ACAP > 4$, showing strong analytical expertise on the team.
    These thresholds reveal not only how the model makes decisions, but also why certain configurations succeed, which allows stakeholders to reason through and improve suboptimal cases with concrete, domain-aligned actions.
\end{enumerate}
In summary:
\begin{quote}
{\bf  Answering RQ2:  Comparison to standard XAI.} {\em EZR provides explanations that go beyond attribution-based methods like LIME, SHAP, or BreakDown. By mapping instances to concrete cohorts with summary statistics, it delivers stable, rule-based, and actionable insights rather than abstract contributions. Within Pearl’s Ladder of Causation, EZR spans all three rungs(associations through global feature importance, interventions through threshold-based what-if analysis, and counterfactuals through alternative leaf assignments) whereas attribution methods remain confined to associations. Consistent with our \textbf{Maximum Clarity Heuristic}, EZR achieves this explanatory power while relying on fewer, more informative features. In doing so, EZR aligns closely with practitioner triggers (Table~\ref{tab:xai_triggers}), offering explanations that are interpretable, consistent, and directly useful in real-world SE tasks.}
\end{quote}

\subsubsection{Explanations in Action: What the Features Reveal}
In order to test whether explanations aid optimization in practice, we treat each method’s feature importance scores as a feature selector and measure downstream utility. Inspired by XAI’s actionability principle, we convert scores into a compact subset, train an independent learner on those features, and compare against all-features and common SE feature ranking tools (SHAP, ReliefF, ANOVA). The idea is simple. If an explanation is faithful, its subset should help learners achieve lower the \textit{d2h} or higher win rates than baselines. This setup makes explanation quality a falsifiable claim about utility beyond the method that produced it.

Table \ref{tab:fs-results} evaluates whether explanation-driven feature subsets improve downstream optimization. We compare three commonly used methods in the SE domain, ReliefF, SHAP, ANOVA, alongside using all features, and EZR. Each of the methods is paired with the LightGBM regressor across all datasets\footnote{We have also executed this experiment using other regressors. Refer to this link: \url{https://github.com/amiiralii/Minimal-Data-Maximum-Clarity/tree/main/paper_materials}}. Cells report the median \textit{win(d2h)} for the feature selector and dataset. The $X$ column gives the total number of features in the dataset; $X^*$ is the number of features selected(for treatment "all" $X=X^*$). “Budget” records how many labels EZR used to compute feature importances. (It is either 150 or 40\% of the training set, whichever is smaller) We have also used color coding to aid comparison. White marks the per-dataset best(there is at least 1 white cell in each row); Blue reaches $\ge90\%$ of the referenced optimal; Yellow and red are  $\ge75\%$ and  $\leq75\%$ of the referenced optimal, respectively. Within a given regressor block, read \emph{across a row} to compare feature selectors for the same dataset. EZR should be interpreted jointly with the Budget. Cases where small subsets trained from limited labels still land in the white/blue region indicate explanations that capture optimization-relevant signal.

\begin{table}[hbpt]
\scriptsize
\centering
\resizebox{4.5in}{!}{%
{\scriptsize \begin{tabular}{
>{\columncolor[HTML]{FFFFFF}}c 
>{\columncolor[HTML]{FFFFFF}}c 
>{\columncolor[HTML]{FFFFFF}}c 
>{\columncolor[HTML]{FFFFFF}}c 
>{\columncolor[HTML]{FFFFFF}}c |c|ccc|cc}
\toprule
\textbf{RLF} & \textbf{SHAP} & \textbf{EZR} & \textbf{anova} & \textbf{all} & \textbf{data} & \textbf{x*} & \textbf{x} & \textbf{y} & \textbf{budget} & \textbf{rows} \\
\midrule
93 & 93 & 93 & 93 & 93 & SS-A & 3 & 3 & 2 & 150 & 1344 \\
84 & 84 & 84 & 84 & 84 & SS-B & 2 & 3 & 2 & 80 & 207 \\
84 & 84 & 84 & 84 & 84 & SS-C & 3 & 3 & 2 & 150 & 1513 \\
77 & 77 & 77 & 77 & 77 & SS-D & 3 & 3 & 2 & 75 & 197 \\
97 & 97 & 97 & 97 & 97 & SS-E & 3 & 3 & 2 & 150 & 757 \\
98 & 98 & 98 & \cellcolor[HTML]{CFE2F3}97 & 98 & SS-F & 2 & 3 & 2 & 75 & 197 \\
93 & \cellcolor[HTML]{CFE2F3}91 & \cellcolor[HTML]{CFE2F3}91 & \cellcolor[HTML]{CFE2F3}91 & 93 & SS-G & 2 & 3 & 2 & 75 & 197 \\
\cellcolor[HTML]{FFE599}80 & 99 & 99 & 99 & 99 & wc-1 & 2 & 3 & 1 & 75 & 197 \\
\cellcolor[HTML]{CFE2F3}94 & 99 & 99 & 99 & 99 & wc-2 & 2 & 3 & 1 & 75 & 197 \\
\cellcolor[HTML]{CFE2F3}94 & 99 & 99 & 99 & 99 & wc-3 & 2 & 3 & 1 & 75 & 197 \\
100 & 100 & 100 & 100 & 100 & SS-H & 2 & 4 & 2 & 100 & 260 \\
99 & 99 & 99 & 99 & 99 & SS-I & 5 & 5 & 2 & 150 & 1081 \\
80 & 80 & 80 & \cellcolor[HTML]{CFE2F3}79 & \cellcolor[HTML]{CFE2F3}79 & auto93 & 3 & 5 & 3 & 150 & 399 \\
66 & 66 & 66 & 66 & 66 & HCI-hard & 5 & 5 & 3 & 150 & 10001 \\
100 & 100 & 100 & 100 & 100 & HCI-easy & 4 & 5 & 3 & 150 & 10001 \\
\midrule
96 & \cellcolor[HTML]{CFE2F3}95 & 96 & \cellcolor[HTML]{CFE2F3}95 & 96 & SS-J & 5 & 6 & 2 & 150 & 3841 \\
96 & 96 & 96 & 96 & 96 & SS-K & 5 & 6 & 2 & 150 & 2881 \\
100 & 100 & 100 & 100 & 100 & SS-S & 5 & 6 & 2 & 150 & 3841 \\
100 & 100 & 100 & 100 & 100 & rs-1 & 5 & 6 & 1 & 150 & 3841 \\
100 & 100 & 100 & 100 & 100 & rs-2 & 5 & 6 & 1 & 150 & 3841 \\
94 & 94 & 94 & 94 & 94 & sol-1 & 5 & 6 & 1 & 150 & 2867 \\
100 & 100 & 100 & 100 & 100 & wc-4 & 4 & 6 & 1 & 150 & 2881 \\
95 & 95 & 95 & 95 & 95 & Apache-AM & 4 & 9 & 1 & 75 & 193 \\
\cellcolor[HTML]{CFE2F3}87 & 90 & \cellcolor[HTML]{CFE2F3}88 & 90 & \cellcolor[HTML]{CFE2F3}88 & pom3a & 7 & 9 & 3 & 150 & 20001 \\
\cellcolor[HTML]{CFE2F3}85 & 87 & 87 & \cellcolor[HTML]{CFE2F3}85 & 87 & pom3b & 6 & 9 & 3 & 150 & 20001 \\
73 & \cellcolor[HTML]{CFE2F3}68 & \cellcolor[HTML]{CFE2F3}71 & 73 & 73 & pom3c & 7 & 9 & 3 & 150 & 20001 \\
88 & 88 & \cellcolor[HTML]{FFE599}76 & \cellcolor[HTML]{CFE2F3}85 & \cellcolor[HTML]{CFE2F3}82 & pom3d & 7 & 9 & 3 & 150 & 501 \\
\cellcolor[HTML]{CFE2F3}77 & \cellcolor[HTML]{CFE2F3}77 & 80 & \cellcolor[HTML]{CFE2F3}77 & \cellcolor[HTML]{CFE2F3}77 & Wine & 6 & 10 & 2 & 150 & 1600 \\
\cellcolor[HTML]{CFE2F3}98 & 99 & 99 & 99 & 99 & SS-L & 6 & 11 & 2 & 150 & 1024 \\
\cellcolor[HTML]{CFE2F3}91 & 99 & \cellcolor[HTML]{CFE2F3}96 & 99 & \cellcolor[HTML]{CFE2F3}96 & SS-O & 6 & 11 & 2 & 150 & 973 \\
\cellcolor[HTML]{CFE2F3}98 & 99 & \cellcolor[HTML]{CFE2F3}98 & 99 & 99 & SS-P & 6 & 11 & 2 & 150 & 1024 \\
\cellcolor[HTML]{DD7E6B}60 & \cellcolor[HTML]{DD7E6B}59 & \cellcolor[HTML]{DD7E6B}41 & \cellcolor[HTML]{DD7E6B}59 & 97 & SS-X & 1 & 11 & 2 & 150 & 86059 \\
\cellcolor[HTML]{FFE599}86 & \cellcolor[HTML]{CFE2F3}99 & 100 & \cellcolor[HTML]{CFE2F3}99 & 100 & SS-T & 5 & 12 & 2 & 150 & 5185 \\
\cellcolor[HTML]{CFE2F3}93 & \cellcolor[HTML]{CFE2F3}97 & \cellcolor[HTML]{CFE2F3}97 & \cellcolor[HTML]{CFE2F3}97 & 98 & SS-Q & 7 & 13 & 3 & 150 & 2737 \\
\midrule
\cellcolor[HTML]{CFE2F3}95 & 100 & 100 & 100 & 100 & HSMGP & 6 & 14 & 1 & 150 & 3457 \\
87 & \cellcolor[HTML]{CFE2F3}81 & \cellcolor[HTML]{CFE2F3}82 & \cellcolor[HTML]{CFE2F3}82 & \cellcolor[HTML]{CFE2F3}85 & SS-R & 7 & 14 & 2 & 150 & 3009 \\
\cellcolor[HTML]{CFE2F3}94 & \cellcolor[HTML]{CFE2F3}95 & \cellcolor[HTML]{DD7E6B}63 & \cellcolor[HTML]{DD7E6B}57 & 96 & SS-V & 4 & 16 & 2 & 150 & 6841 \\
\cellcolor[HTML]{DD7E6B}74 & \cellcolor[HTML]{DD7E6B}74 & \cellcolor[HTML]{DD7E6B}59 & \cellcolor[HTML]{DD7E6B}74 & 99 & SS-W & 7 & 16 & 2 & 150 & 65537 \\
100 & 100 & 100 & 100 & 100 & X264-AM & 6 & 16 & 1 & 150 & 1153 \\
100 & 100 & 100 & \cellcolor[HTML]{CFE2F3}99 & 100 & SS-M & 7 & 17 & 3 & 150 & 865 \\
81 & 81 & 81 & 81 & 81 & SS-N & 8 & 17 & 2 & 150 & 53663 \\
\cellcolor[HTML]{CFE2F3}59 & 62 & \cellcolor[HTML]{FFE599}53 & \cellcolor[HTML]{FFE599}52 & \cellcolor[HTML]{CFE2F3}58 & coc1000 & 9 & 20 & 5 & 150 & 1001 \\
\cellcolor[HTML]{CFE2F3}95 & \cellcolor[HTML]{CFE2F3}96 & \cellcolor[HTML]{CFE2F3}96 & \cellcolor[HTML]{CFE2F3}93 & 100 & SS-U & 7 & 21 & 2 & 150 & 4609 \\
98 & 98 & \cellcolor[HTML]{CFE2F3}93 & 98 & 98 & xomo\_flight & 7 & 27 & 4 & 150 & 10001 \\
92 & 92 & \cellcolor[HTML]{CFE2F3}90 & \cellcolor[HTML]{CFE2F3}90 & 92 & xomo\_grnd & 7 & 27 & 4 & 150 & 10001 \\
\cellcolor[HTML]{CFE2F3}94 & 95 & 95 & 95 & 95 & xomo\_osp & 7 & 27 & 4 & 150 & 10001 \\
\cellcolor[HTML]{CFE2F3}87 & 88 & \cellcolor[HTML]{CFE2F3}82 & \cellcolor[HTML]{CFE2F3}82 & 88 & xomo\_osp2 & 8 & 27 & 4 & 150 & 10001 \\
\cellcolor[HTML]{CFE2F3}77 & \cellcolor[HTML]{FFE599}75 & \cellcolor[HTML]{FFE599}71 & \cellcolor[HTML]{FFE599}64 & 85 & SQL-AM & 10 & 39 & 1 & 150 & 4654 \\
\cellcolor[HTML]{CFE2F3}87 & \cellcolor[HTML]{CFE2F3}91 & \cellcolor[HTML]{CFE2F3}92 & \cellcolor[HTML]{FFE599}74 & 94 & billing10k & 7 & 88 & 3 & 150 & 10001 \\
\cellcolor[HTML]{FFE599}75 & \cellcolor[HTML]{FFE599}75 & \cellcolor[HTML]{FFE599}81 & \cellcolor[HTML]{FFE599}75 & 97 & Scrum100k & 7 & 124 & 3 & 150 & 100001 \\
\cellcolor[HTML]{FFE599}82 & \cellcolor[HTML]{CFE2F3}88 & \cellcolor[HTML]{CFE2F3}92 & \cellcolor[HTML]{CFE2F3}88 & 97 & Scrum10k & 8 & 124 & 3 & 150 & 10001 \\
\cellcolor[HTML]{CFE2F3}83 & 89 & \cellcolor[HTML]{CFE2F3}86 & \cellcolor[HTML]{CFE2F3}83 & 89 & Scrum1k & 7 & 124 & 3 & 150 & 1001 \\
\cellcolor[HTML]{FFE599}81 & 99 & \cellcolor[HTML]{CFE2F3}93 & \cellcolor[HTML]{FFE599}81 & 99 & FFM-125 & 9 & 128 & 3 & 150 & 10001 \\
\cellcolor[HTML]{FFE599}81 & 97 & \cellcolor[HTML]{CFE2F3}89 & 97 & 97 & FFM-250 & 8 & 256 & 3 & 150 & 10001 \\
95 & \cellcolor[HTML]{CFE2F3}93 & \cellcolor[HTML]{CFE2F3}93 & 95 & 95 & FFM-500 & 8 & 510 & 3 & 150 & 10001 \\
\cellcolor[HTML]{FFE599}81 & 96 & \cellcolor[HTML]{CFE2F3}87 & 96 & 96 & FM-500-1 & 8 & 511 & 3 & 150 & 10001 \\
\cellcolor[HTML]{FFE599}86 & \cellcolor[HTML]{CFE2F3}91 & \cellcolor[HTML]{CFE2F3}93 & \cellcolor[HTML]{CFE2F3}91 & 97 & FM-500-2 & 7 & 511 & 3 & 150 & 10001 \\
95 & 95 & 95 & \cellcolor[HTML]{CFE2F3}87 & 95 & FM-500-3 & 9 & 513 & 3 & 150 & 10001 \\
97 & 97 & \cellcolor[HTML]{CFE2F3}93 & \cellcolor[HTML]{CFE2F3}96 & 97 & FM-500-4 & 9 & 517 & 3 & 150 & 10001 \\
\cellcolor[HTML]{FFE599}82 & \cellcolor[HTML]{CFE2F3}94 & \cellcolor[HTML]{CFE2F3}94 & \cellcolor[HTML]{DD7E6B}71 & 100 & FFM-1000 & 7 & 1044 & 3 & 150 & 10001 \\
\bottomrule
\end{tabular}
}}
\caption{EZR Performance as a feature selection tool. White cells show the best performance; blue and yellow cells mark EZR answers that achieve at least 90\% and 75\% of the best; }
\label{tab:fs-results}
\end{table}

\begin{figure}[htbp]
  \centering
  \includegraphics[width=\textwidth]{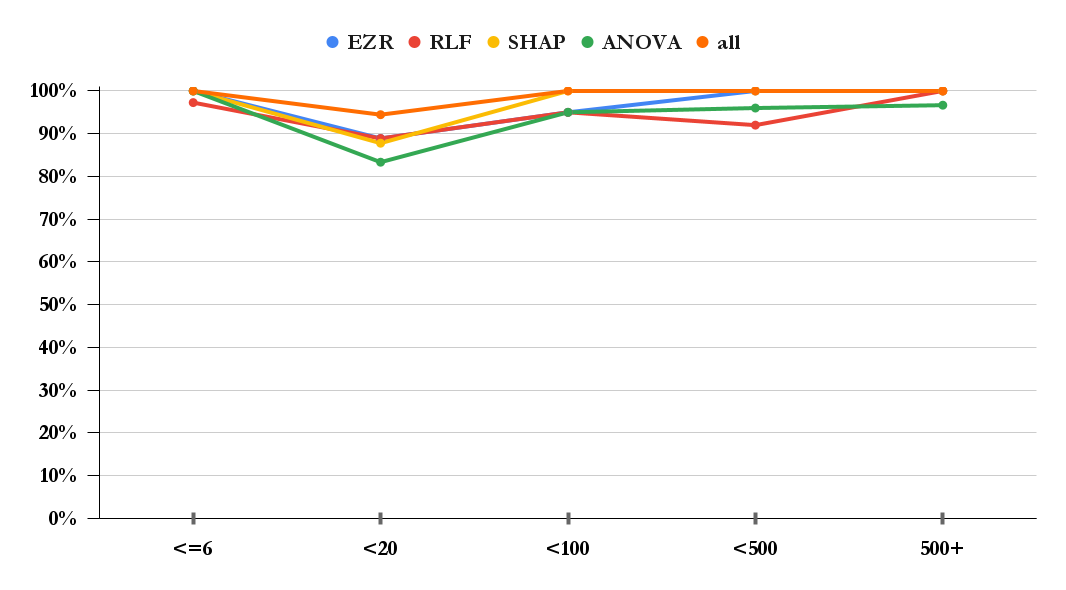}
  \caption{In what portion of each dataset group does each method achieve 75\% of the referenced optimal or closer?}
  \label{fig:summary_75}
\end{figure}

\begin{figure}[htbp]
  \centering
  \includegraphics[width=\textwidth]{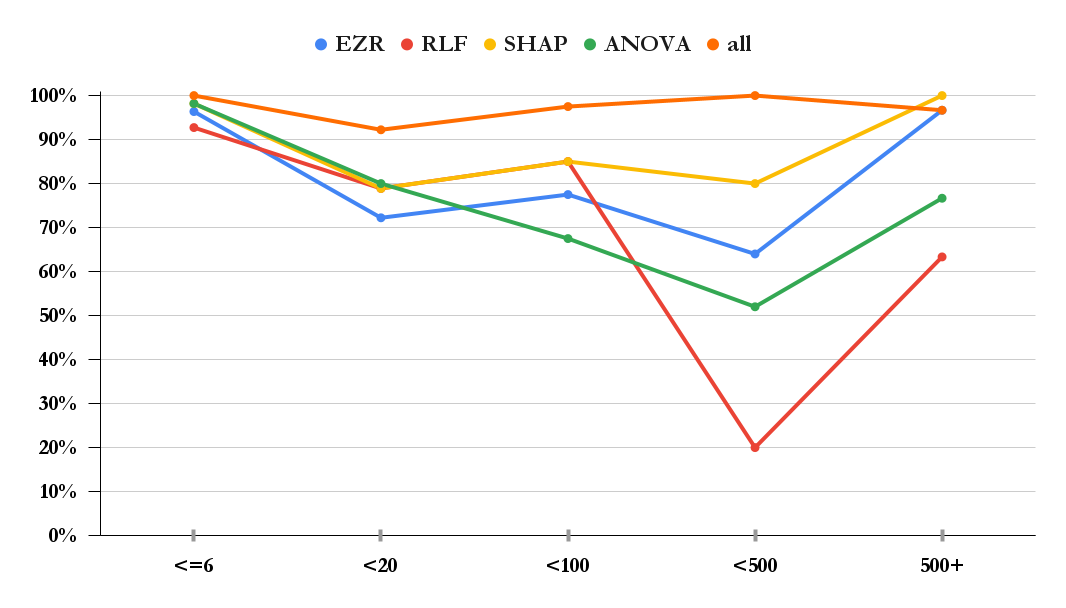}
  \caption{In what portion of each dataset group does each method achieve 90\% of the referenced optimal or closer?}
  \label{fig:summary_90}
\end{figure}

Figures~\ref{fig:summary_75} and~\ref{fig:summary_90} present an aggregated view of Table~\ref{tab:fs-results}, summarizing outcomes across all treatments. To evaluate feature-reduction efficiency, datasets were grouped into five bins based on feature count($\leq$6, $<$20, $<$100, $<$500, and $500+$). With 60 datasets and 5 regressors, this results in 300 treatments distributed across these bins. Each point in the figures shows the proportion of treatments per each method in a given bin that achieved at least 75\% (Figure~\ref{fig:summary_75}) or 90\% (Figure~\ref{fig:summary_90}) of the referenced optimal performance. 

Overall, the aggregated results reveal two key trends. In the 75\% threshold analysis (Figure~\ref{fig:summary_75}), the performance of all methods largely converges, with little separation across buckets. This “blurring” effect indicates that EZR is just as effective as other approaches while relying on far fewer labels. In contrast, the 90\% threshold analysis (Figure~\ref{fig:summary_90}) shows clearer differences, particularly in the larger-feature buckets, where all methods fall short. The absence of such gaps at the 75\% level suggests that a substantial portion of results lies between 75–90\% of optimal performance. Importantly, only SHAP consistently outperforms EZR in these higher thresholds, yet it does so with full supervision, whereas EZR achieves comparable performance using only 150 labels. 

In summary: 

\begin{quote}
{\bf  Answering RQ3: Practical utility of explanations.} {\em EZR’s explanation-driven feature rankings substantially improve downstream optimization, performing on par with or better than established feature selection methods under most conditions. Although SHAP occasionally yields stronger results in high-dimensional settings, EZR achieves comparable performance while relying on far fewer labels. This outcome reinforces the Maximum Clarity Heuristic: in complex software engineering tasks, effective explanations and strong optimization performance can be achieved with less data, not more. }
\end{quote}

\section{Discussion}

Our results suggest that EZR’s remarkable label efficiency and robust performance stem from two core mechanisms: selective sampling that skips noisy or irrelevant configurations, and the use of a compact decision tree to guide further exploration. By focusing only on those samples most likely to yield improvements, EZR avoids the “needle‐in‐a‐haystack” problem that plagues exhaustive search, and then leverages a small decision tree to rapidly identify promising directions.

\subsection{Why EZR Works}
\begin{enumerate} [noitemsep, topsep=0pt]
  \item \textbf{Noise avoidance and relevance filtering.}  
    Rather than uniformly sampling the entire configuration space, EZR’s warm‐start and active learning steps label the “corners” of the configuration space, the extreme parameter combinations that are most likely to lie on the Pareto front. By doing so, it avoids wasting budget on noisy or redundant regions and concentrates sampling where performance variation is highest. This strategy is well‐suited to software engineering data, which often resides on low‐dimensional manifolds. Lustosa and Menzies’s Dimensionality Reduction Ratio (DRR) study shows that a small fraction of intrinsic dimensions captures nearly all the variability in software engineering performance metrics \cite{ref20}.

  \item \textbf{Lightweight surrogate modeling.}  
    The decision tree induced from these select labels is necessarily small. Its depth and branching factor are both bounded by the number of samples. Such a tree can be traversed in $O(\log n)$ time to suggest tweaks that push toward better performance, and its human‐readable structure offers immediate interpretability.

  \item \textbf{Sensitivity to labeling budget.} To assess how sensitive EZR is to the chosen labeling budget, we repeated the same optimization protocol under alternative budgets and compared the resulting performance distributions to our default setting. For \textit{light} and \textit{medium} datasets, we varied the absolute labeling budget $\{10,25,50,100,200\}$ and for \textit{heavy} datasets, we varied the budget as a fraction of the training set $\{1\%,5\%,10\%,20\%,40\%\}$. For each dataset and each budget treatment, we repeated optimization 20 times using different random seeds, producing a distribution of $\textit{win}$ scores per treatment. We then compared each treatment’s $\textit{win}$ distribution against the distribution obtained with our selected budget (50 labels for light/medium and 10\% for heavy), and counted how many datasets (out of 60) were \emph{significantly different} from the selected budget. The resulting counts are summarized in Figure~\ref{fig:budget-small} (light/medium) and Figure~\ref{fig:budget-heavy} (heavy). Overall, the sensitivity analysis shows a clear pattern. Budgets below the selected defaults are frequently and significantly worse, whereas increasing the budget beyond the selected defaults rarely produces statistically distinguishable improvements.

\begin{figure}[t]

  \centering
  \begin{minipage}{0.49\linewidth}
    \centering
  \fbox{\includegraphics[width=\textwidth]{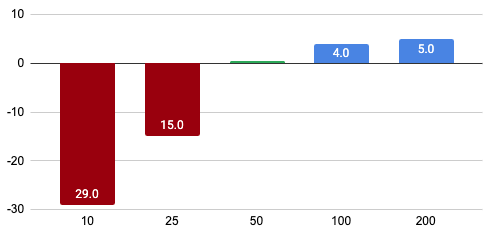}}
    \caption{Sensitivity on light and medium datasets (41 in total). Red indicates the number of datasets performing significantly weaker than our chosen budget, if we use too few labels (fewer than 50). The blue columns show the number of datasets performing significantly better than our chosen budget (using higher budgets), offering very little additional benefit.}
    \label{fig:budget-small}
  \end{minipage}\hfill
  \begin{minipage}{0.5\linewidth}
    \centering
  \fbox{\includegraphics[width=\textwidth]{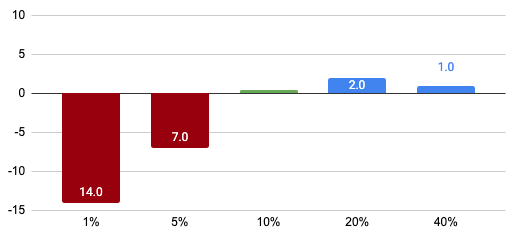}}
    \caption{Sensitivity on heavy datasets (19 in total). Red indicates the number of datasets performing significantly weaker than our chosen budget, if we use too few labels (fewer than 50). The blue columns show the number of datasets performing significantly better than our chosen budget (using higher budgets), offering very little additional benefit.}
    \label{fig:budget-heavy}
  \end{minipage}
\end{figure}

\color{black}

\end{enumerate}

\subsection{Limitations and Caveats}
\begin{itemize}[noitemsep, topsep=0pt]
  \item \textbf{Dataset Characteristics:}  
    EZR’s efficiency hinges on the idea that high‐quality configurations can be (at least partially) ordered by their performance. In problems where the objective landscape is highly irregular or multimodal with many local optima that offer conflicting trade-offs, the sampling strategy may fail to isolate the true Pareto front, leading to suboptimal recommendations. Moreover, because EZR focuses on the most promising areas, it may systematically overlook rare but critical configurations that lie outside the main performance clusters. In safety-critical contexts, explicit mechanisms to ensure coverage of edge-case scenarios may be necessary.

  \item \textbf{Problem setting:}  
    As the number of available parameter values grows, the size and complexity of the induced decision tree can rise, demanding more labels to maintain predictive accuracy. More complex feature configuration spaces may require additional strategies to keep EZR efficient. Also, EZR assumes a fixed configuration space and a single-shot labeling process. In dynamic or streaming environments, where system behavior drifts over time or new options appear, EZR would need extensions for incremental learning and adaptive re-budgeting to remain effective.

\end{itemize}

By highlighting both the theoretical rationale and practical boundaries of EZR, we provide practitioners with a clear understanding of when and how this lightweight, interpretable heuristic can be most effectively applied and when more exhaustive or specialized methods remain necessary.

\subsection{Threats to validity}
To ensure rigor and transparency, we’ve examined key factors that might affect our results and rerun experiments under varied settings to confirm their stability. Any conclusions should be interpreted with the following potential limitations in mind:
\begin{enumerate}[noitemsep, topsep=0pt]
    \item \textbf{Conclusion validity}: Conclusion validity threats arise when the observed effects may be due to random variation or the choice of specific analysis procedures rather than reflecting true performance differences. EZR introduces two sources of randomness. The initial warm-start selection of rows and the sampling performed within the active learning loop. To reduce the influence of any single random sequence, we repeated every experiment 20 times, each with a unique random seed, and applied a fresh 80/20 train/test split for each run. Reported results are the mean over these 20 trials, ensuring that our findings are robust to stochastic variation and not artifacts of chance.

    \item \textbf{Internal Validity}: Internal validity threats concern whether the performance differences observed are attributable to our method rather than to uncontrolled factors. We minimized this risk by employing consistent experimental protocols across all treatments. Identical dataset splits, identical evaluation metrics, and fixed labeling budgets for different treatments. All baseline models were implemented using well-established libraries with default or widely accepted hyperparameters, ensuring fair and reproducible comparison.

    A deliberate difference between EZR and the baselines is label availability, as label efficiency is central to our research questions. In the RQ1 experiment, for light and medium datasets, EZR’s labeling budget was fixed at 60 examples, derived from performing 50 active learning iterations to expand the decision tree, followed by selecting the top 10 candidates from the unlabeled pool for final evaluation. This choice of 50 iterations follows the finding by Ganguly and Menzies \cite{ref16} that most datasets reach state-of-the-art performance at this point. For heavy datasets, which embed more complex patterns, we evaluated budgets of 1\%, 5\%, 10\%, 15\%, and 20\% of the training data. Performance gains beyond 10\% were negligible. Accordingly, we fixed the budget for heavy datasets at 10\% of the training set.

    Selecting representative baselines is also critical to avoid biased comparisons. We chose one strong performer from each major regression family. Linear regression, decision trees, bagging (Random Forests), boosting ensembles (LightGBM), neural networks, and support vector machines, ensuring that our predictive model comparisons span both simple and highly optimized learners. For feature ranking, we included SHAPley values (an explanation-driven approach), ReliefF (a high-performing filter method), ANOVA (a classic statistical test), and an all-features baseline to capture exhaustive selection. While no single baseline consistently outperforms all others across every dataset, this diversity provides a broad and rigorous context against which to evaluate EZR.
    
    \item \textbf{Construct validity}: Construct validity threats arise when the chosen evaluation measures do not accurately capture the intended study objectives. We assess optimization performance using the \textit{d2h} score, from which we derive a “win rate” across methods. To verify that our conclusions are not dependent on this specific metric, we repeated all experiments using the Chebyshev distance \cite{ref14,ref15}. While absolute scores differed, the relative rankings and overall conclusions remained stable.

    In addition, because statistical rank-based tests such as Scott Knott do not convey how close methods are in absolute terms, we introduced a relative performance metric. This metric quantifies how near EZR comes to the best-known solution achieved by a fully supervised state-of-the-art method, thereby directly reflecting the trade-off between accuracy and labeling cost.

    For feature selection, we evaluated predictive fidelity within the optimization task, acknowledging that these results may not transfer directly to other tasks such as classification or regression.
    
    \item \textbf{External validity}: External validity threats concern the extent to which our findings generalize beyond the studied datasets. Real-world software engineering optimization problems vary widely in feature distributions, noise characteristics, and cost structures, which small benchmark collections may fail to capture. To address this, we evaluated EZR on 60 diverse datasets(far exceeding the dataset counts used in prior work) from the MOOT repository, which covers configuration, hyperparameter tuning, and process optimization scenarios. The MOOT repository was specifically developed to enhance the generality of search-based software engineering conclusions. It aggregates datasets from numerous recent SE and AI publications into a unified testbed that reflects a broad spectrum of real-world optimization challenges. Leveraging MOOT, therefore, strengthens the external validity of our study and increases the likelihood that our conclusions apply to other problem domains.

\end{enumerate}

\subsection{Future Work}

A promising direction is to move beyond fixed labeling budgets by developing adaptive schemes that monitor convergence or model uncertainty at each step and automatically allocate or halt sampling when marginal gains wane. Complementing this, we plan to validate EZR on a broader array of software engineering benchmarks drawn from different repositories and use cases (e.g., build‐system tuning, static‐analysis parameterization, continuous‐integration workflows) to test its robustness across practical scenarios. We also see value in embedding EZR as the front‐end of hybrid optimization pipelines, using its rapid “good‐enough” search to identify a promising region before handing off to more exhaustive global optimizers for fine‐tuning. Finally, integrating EZR’s active learning strategy into automated machine learning frameworks, where it could drive both configuration and feature engineering choices, offers a path to end‐to‐end systems that balance interpretability, efficiency, and accuracy.

\section{Conclusion}
We have presented  EZR, a simple yet effective framework for multi-objective optimization and explanation in software engineering. Instead of relying on large labeled datasets and complex models, EZR uses a small, actively selected subset of configurations to build concise decision‐tree explanations and surrogate predictors. Evaluated on over 60 diverse MOOT datasets, spanning configuration, hyperparameter tuning, and process optimization, EZR delivers near‐optimal settings while querying only a fraction of the labels required by state‐of‐the‐art methods.

First, in optimization, based on tables \ref{tab:light-results-1}, \ref{tab:medium-results-1}, and \ref{tab:heavy-results-1}, EZR achieves within 90\% of the best-known performance in 77\% of datasets (and greater than 70\% in 92\% of cases) using only 60 labels (or 10\% of the data for large problems), versus full supervision for baselines. Its simplicity and label efficiency make it an attractive choice for rapid tuning in domains that currently lack systematic optimization tools. In practice, EZR is ideal when a “good‐enough” setting is needed quickly and with minimal effort. However, because EZR does not guarantee the absolute optimum, critical or high-stake applications (e.g., healthcare or finance) may still require more exhaustive or specialized optimizers. 
In addition, we further compared EZR against established optimizers such as SMAC and DEHB. Despite its simplicity and lower labeling cost, EZR achieves competitive performance to these methods across a large fraction of datasets, reinforcing its competitiveness within the broader optimization literature. \color{black}

Second, as a feature ranking tool, according to table \ref{tab:fs-results}, EZR matches or closely approaches the performance of established methods (e.g., SHAP, ReliefF, ANOVA) while using significantly fewer labels. This efficiency makes it ideally suited to scenarios where labeling costs are the primary constraint. At the same time, when marginal gains in predictive accuracy outweigh labeling effort, more aggressive selection techniques may be preferred. Moreover, our results indicate that predictive accuracy hinges more on the choice of regression model and dataset characteristics than on the specific feature ranking method.

Overall, EZR embodies the “less but better” principle. By carefully choosing minimal data, we achieve maximum clarity and strong performance. We hope this work encourages further exploration of lightweight, interpretable heuristics in automated software engineering.

\bibliographystyle{elsarticle-num}
\bibliography{references}

\tracingcommands=1
\end{document}